\documentclass[twoside,onecolumn]{article}

\usepackage{blindtext} 

\usepackage[sc]{mathpazo} 
\usepackage[T1]{fontenc} 
\usepackage{microtype} 

\usepackage[english]{babel} 
\usepackage{graphicx}

\usepackage[hmarginratio=1:1,left=25mm,top=32mm,columnsep=20pt]{geometry} 
\usepackage[hang, small,labelfont=bf,up,textfont=rm,up]{caption} 
\usepackage{booktabs} 

\usepackage{lettrine} 

\usepackage{enumitem} 
\setlist[itemize]{noitemsep} 

\usepackage{abstract} 

\usepackage{titlesec} 
\renewcommand\thesection{\arabic{section}} %
\renewcommand\thesubsection{\thesection.\arabic{subsection}} %
\titleformat{\section}[block]{\large\bf}{\thesection.}{1em}{} 
\titleformat{\subsection}[block]{\centering\it}{\thesubsection.}{1em}{} 

\usepackage{fancyhdr} 
\usepackage{titling} 

\usepackage{hyperref} 

\pretitle{\begin{center}\Large\bfseries} 
\posttitle{\end{center}} 
\title{Hypersonic flow over a wedge in the detached shock range} 
\author{%
\textsc{H. G. Hornung} \\[1ex] 
\normalsize Graduate Aerospace Laboratories, California Institute of Technology\\ 
\normalsize \href{mailto:hans@caltech.edu}{hans@caltech.edu} 
}
\date{} 
\renewcommand{\maketitlehookd}{%
\begin{abstract}
\noindent In a recent publication \cite{Hornung2019} showed
that the shock wave 
stand-off distance and the drag coefficient of a cone in inviscid 
hypersonic flow of a perfect gas
can be expressed as the product of a function of the inverse
normal-shock density ratio $\varepsilon$
and a function of a cone-angle parameter $\eta$, thus reducing the 
number of independent parameters from 
three (Mach number, specific heat ratio and angle) to two. By making a large 
number of Euler computations, analytic forms of the functions were obtained.
In this article the same approach is applied to the symmetrical 
flow over a wedge. It
turns out that the same simplification applies and corresponding analytical
forms of the functions are obtained.  The functions of $\varepsilon$ are
compared with newly determined
corresponding functions for flow over a circular cylinder.
\end{abstract}
}

\begin{document}

\maketitle

\section{Introduction}
One of the most important parameters in hypersonic flow is the inverse
normal-shock density ratio, which for a perfect gas is
\begin{equation}
\varepsilon\,=\,{\rho_\infty\over{\rho_s}}\,=\,{\gamma-1+2/M_\infty^2\over{\gamma+1}},
\end{equation}
where $\rho$ is density, $\gamma$ is the ratio of specific heats, and $M$ is
the Mach number. The subscripts $\infty$ and $s$ refer to the free stream
and to the immediate post-normal-shock condition.

In the flow over a wedge with given free-stream conditions, 
the shock wave is straight and attached to the wedge tip
when the wedge half-angle $\theta$ is sufficiently small. As $\theta$ is increased a point is reached at which the flow downstream of the wedge is sonic,
so that information about the length of the wedge from tip to shoulder
can be communicated to the tip. The shock begins to curve, and, at a very 
slightly larger value of $\theta$, it detaches from the tip. Results of 
Euler computations are shown for these three conditions in Figure~1.
\begin{figure}
   \begin{center}
   \includegraphics[width=0.32\columnwidth]{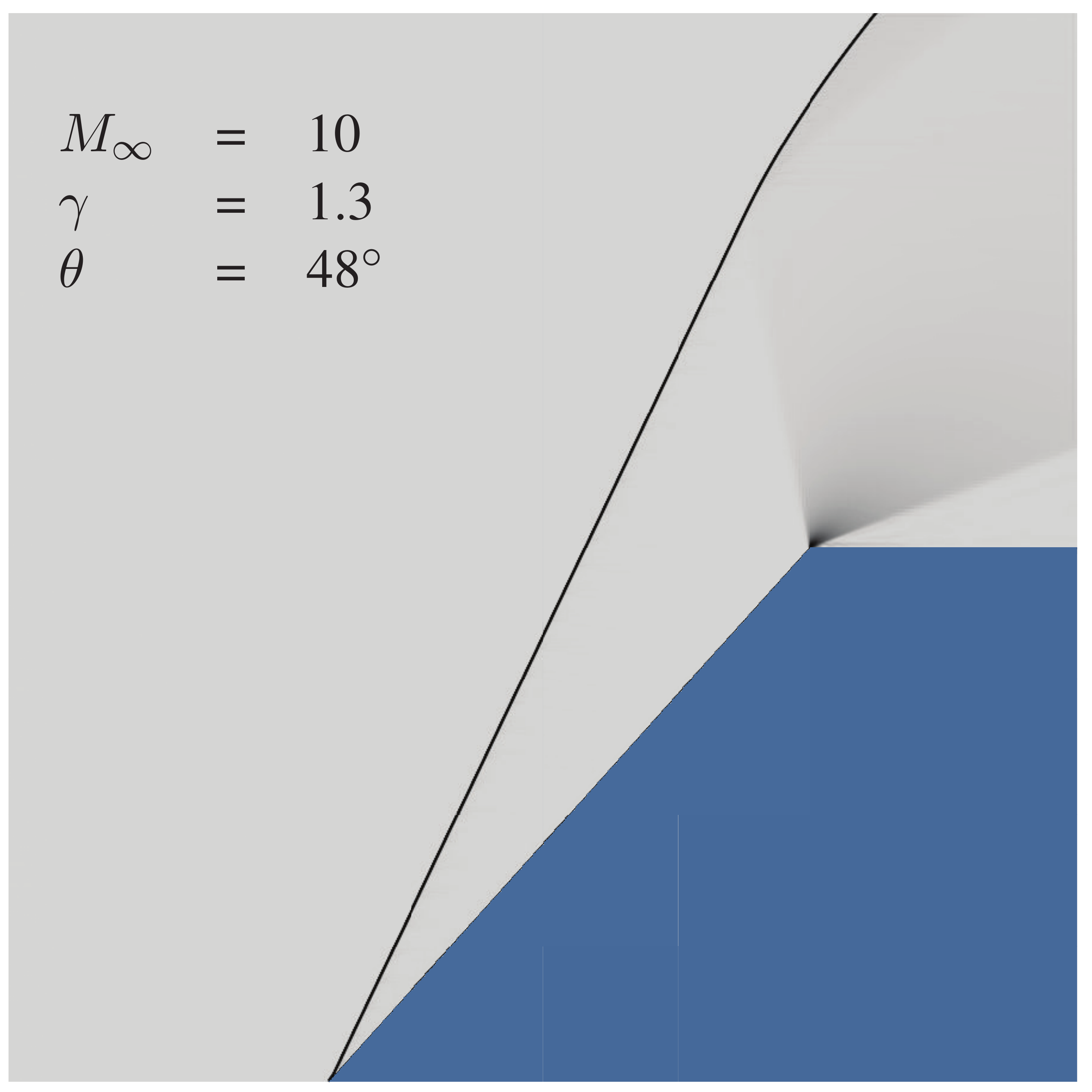} \includegraphics[width=0.32\columnwidth]{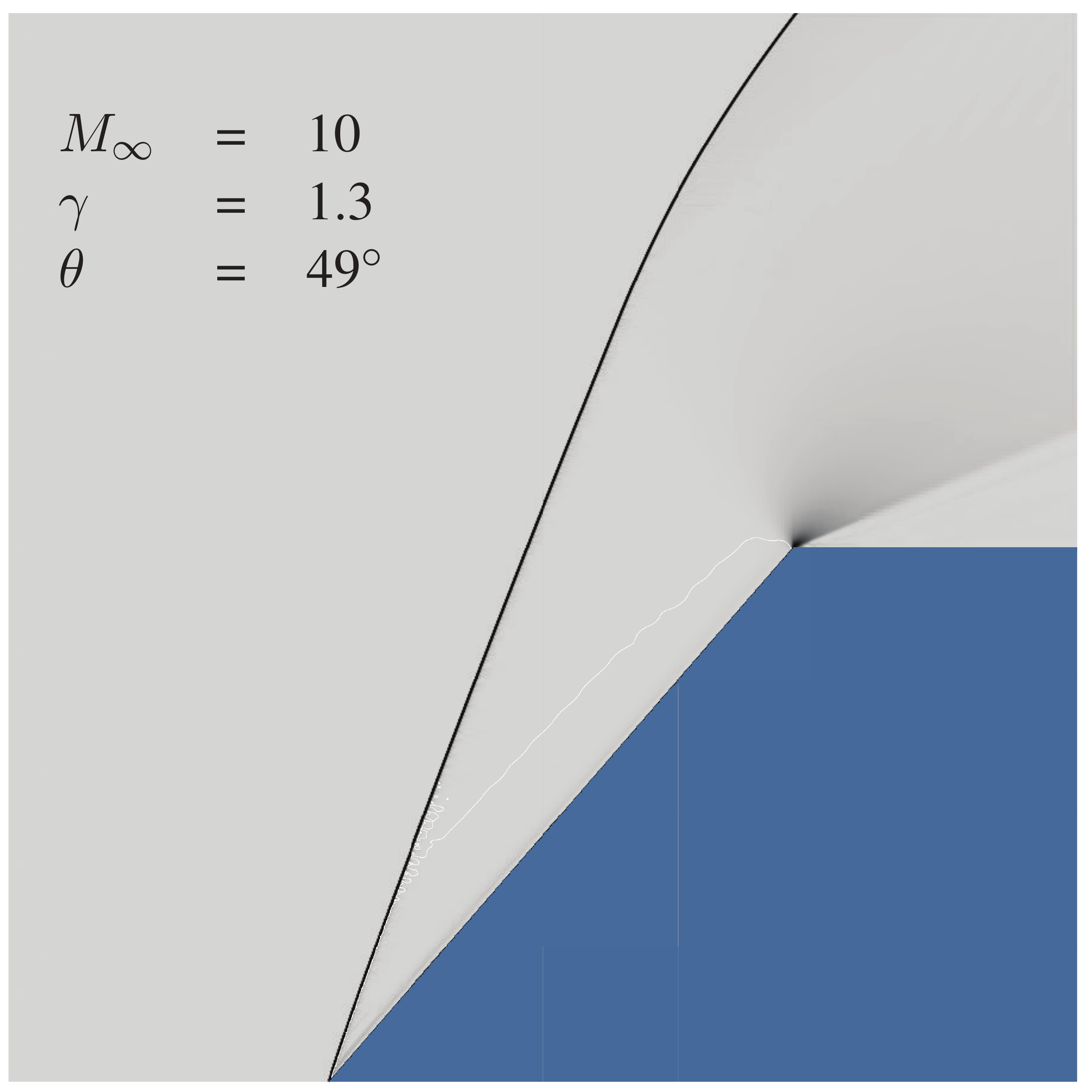} \includegraphics[width=0.32\columnwidth]{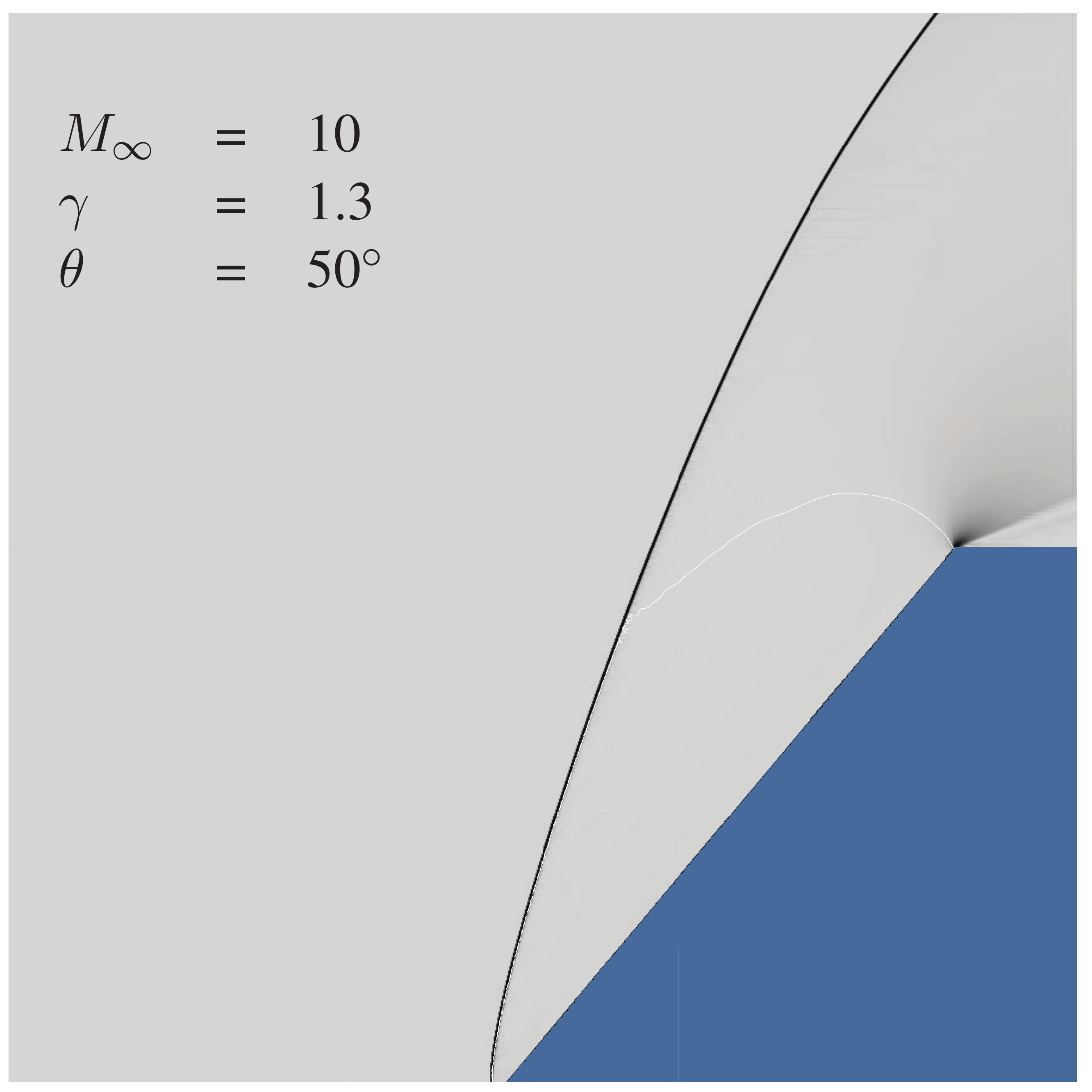}
   \caption{Flow over a wedge at $M_\infty=10$ and $\gamma=1.3$ with
$\theta=48$, 49 and 50$^\circ$, showing (l. to r.) 
entirely supersonic flow with a
straight attached shock, partly subsonic flow with a curved, attached shock and
flow with a detached shock. The white line is the sonic line.}
   \end{center}
\end{figure}
An example of the power of the parameter $\varepsilon$ 
is the approximation given by \cite{Hayes59} for the values of $\theta$ and
the shock angle $\beta$ at detachment.
\begin{equation}
\beta_d\,=\,{\rm arctan}\sqrt{1\over{\varepsilon}},\qquad
\theta_d\,=\,2\beta_d-{\pi\over{2}}.
\end{equation}
Here the subscript $d$ refers to the detachment condition.
The exact values of these detachment angles may be determined from
\begin{equation}
\beta_{de}\,=\,{\rm arcsin}\sqrt{{{(\gamma+1)M_\infty^2/4-1+\sqrt{\gamma+1}\sqrt{(\gamma+1)M_\infty^4/16+1+(\gamma-1)M_\infty^2/2}}\over{\gamma M_\infty^2}}}
\end{equation}
\begin{equation}
\theta_{de}\,=\,{\rm arctan}\left({(M_\infty^2\sin^2\beta_{de}-1)/\tan\beta_{de}\over{1+[(\gamma+1)/2-\sin^2\beta_{de}]M_\infty^2}}\right)
\end{equation}
Figure~2 shows the quality of the approximation of \cite{Hayes59} by plotting
exact values for Mach numbers between 4 and 10 and $\gamma$ between 1.05 and 1.4
together with the approximation vs. $\varepsilon$. As $\varepsilon$ 
increases the approximate value of $\theta_d$ falls above the exact curves,
especially at the lower Mach numbers. 
\begin{figure}[ht!]
   \begin{center}
   \includegraphics[width=0.8\columnwidth]{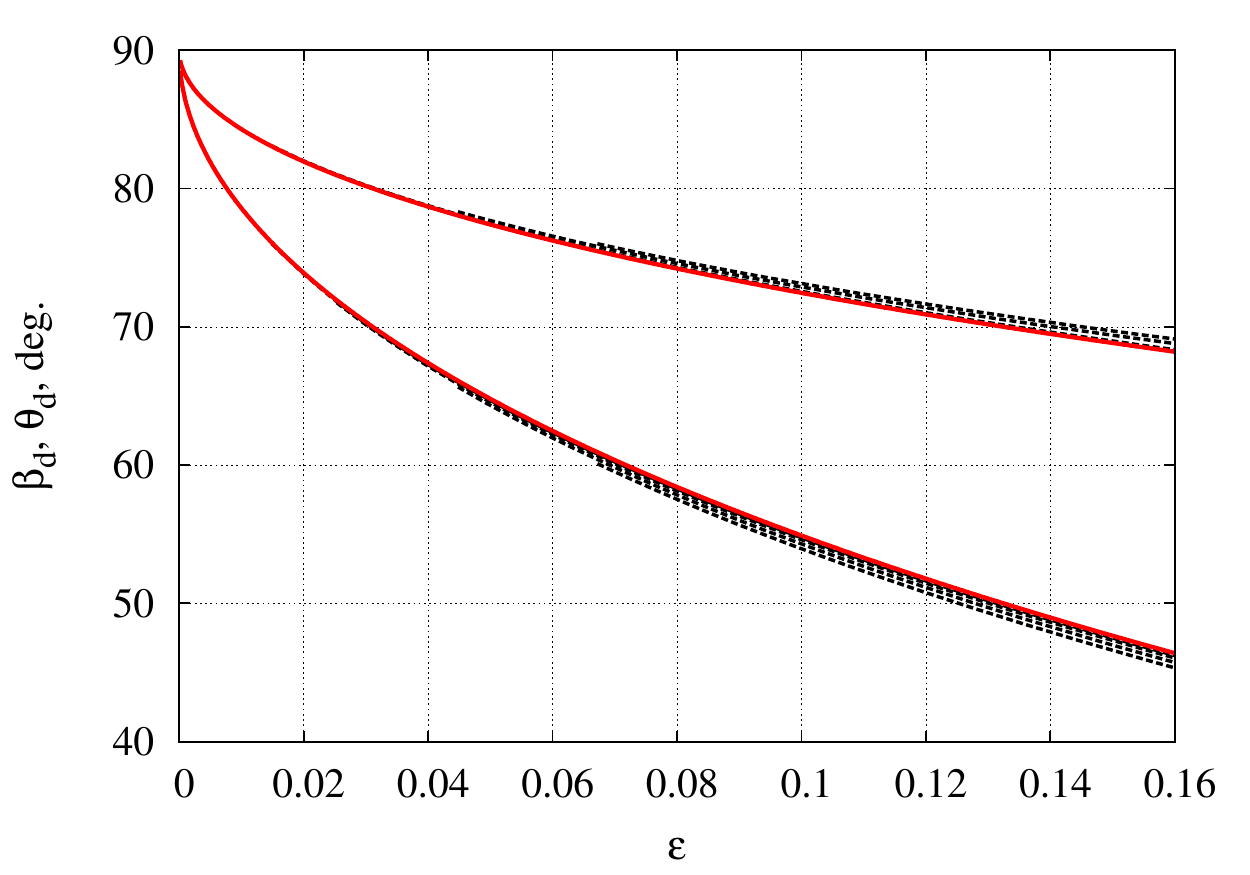}
   \caption{Exact detachment angles for $4<M_\infty<10$ and $1.05<\gamma<1.4$,
      plotted with black dashed lines,
      compared with the approximation of \cite{Hayes59} (equations 3 and 4)
      in red.}
   \end{center}
\end{figure}

The range of $\theta$ for which the shock is detached is of particular interest
here. As in \cite{Hornung2019}, we introduce the variable
\begin{equation}
\eta\,=\,{\theta-\theta_d\over{\pi/2-\theta_d}},
\end{equation}
such that $\eta=0$ at detachment and $\eta=1$ at $\theta=\pi/2$.
Since analytical formulas for the
exact detachment angles exists in the case of wedge flow,  we
use the variable
\begin{equation}
\eta_e\,=\,{\theta-\theta_{de}\over{\pi/2-\theta_{de}}}
\end{equation}
in place of $\eta$.

For the shock stand-off distance, $\Delta$, 
which is of special interest, we again 
make the
hypothesis that it follows the functional form
\begin{equation}
{\Delta\over{H}}\,=\,g(\varepsilon)f(\eta_e)
\end{equation}
as in \cite{Hornung2019}. Here $H$ is the height of the wedge measured
from the symmetry plane to the shoulder.
In order to test the hypothesis, we make a large number of computations
covering the parameter space $(M_\infty, \gamma, \theta)$. If the hypothesis
is true in the case of flow over a wedge as it was for cone flow, 
the results can be used to determine the functional forms of
$g$ and $f$.
\begin{figure}[ht!]
   \begin{center}
   \includegraphics[width=0.24\columnwidth]{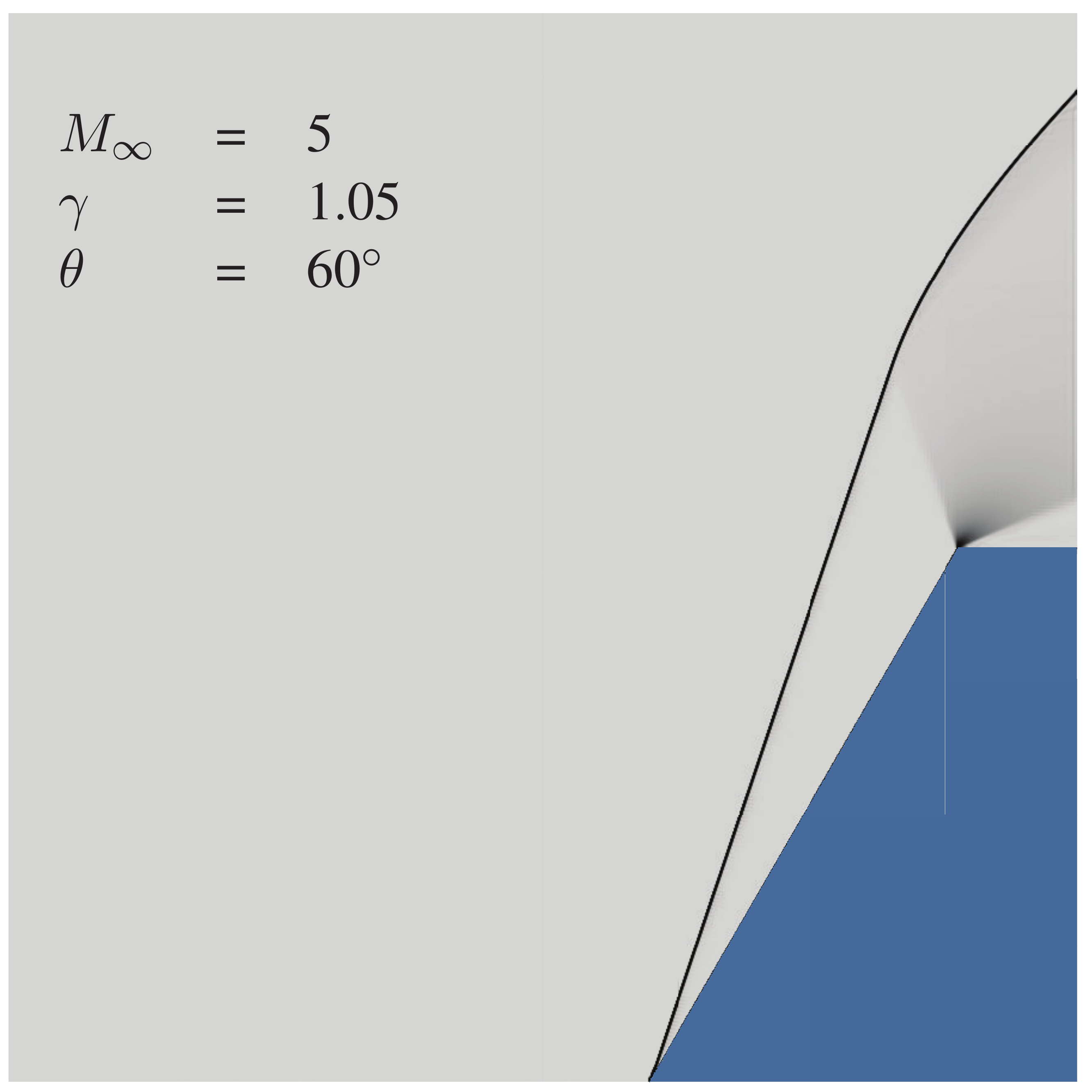} \includegraphics[width=0.24\columnwidth]{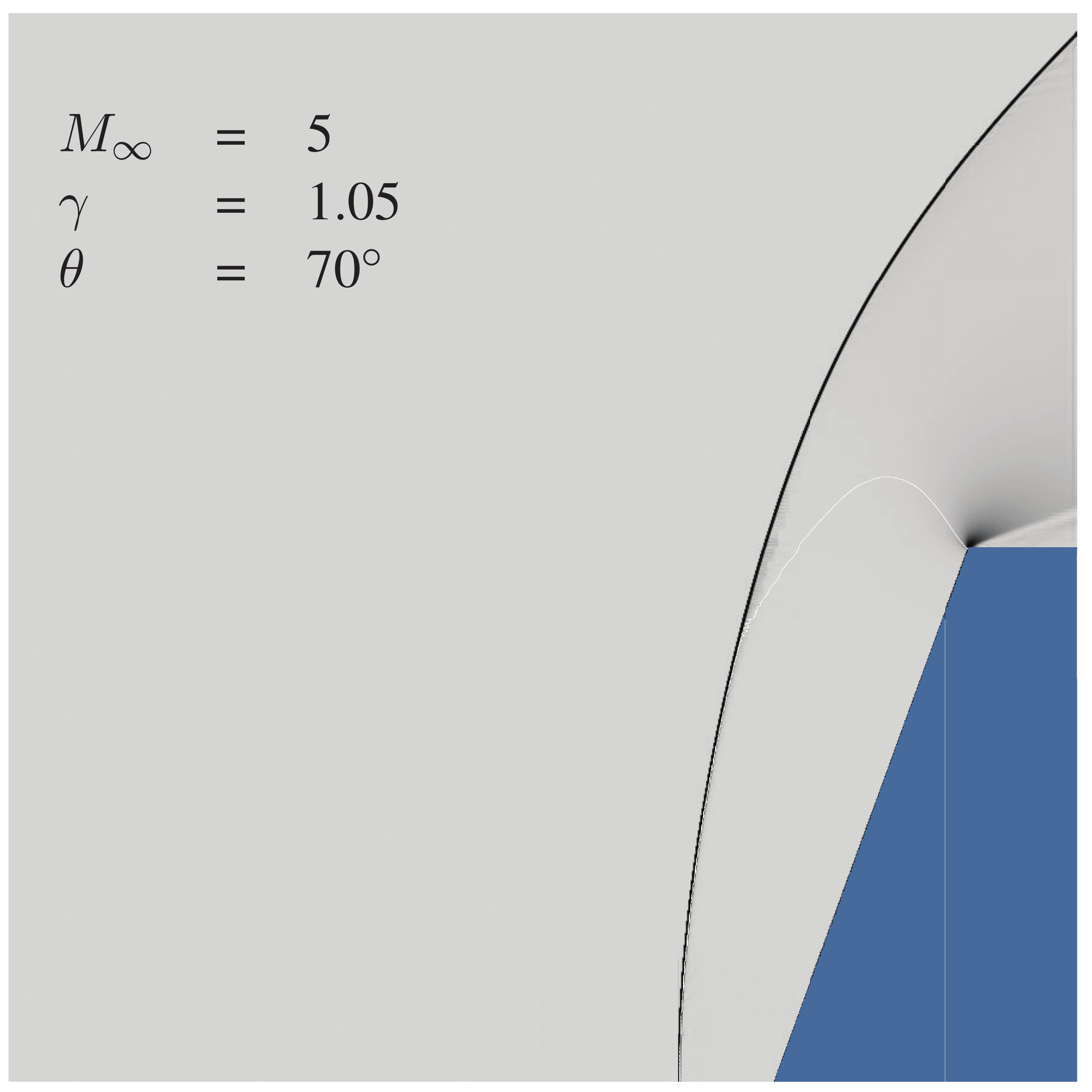} \includegraphics[width=0.24\columnwidth]{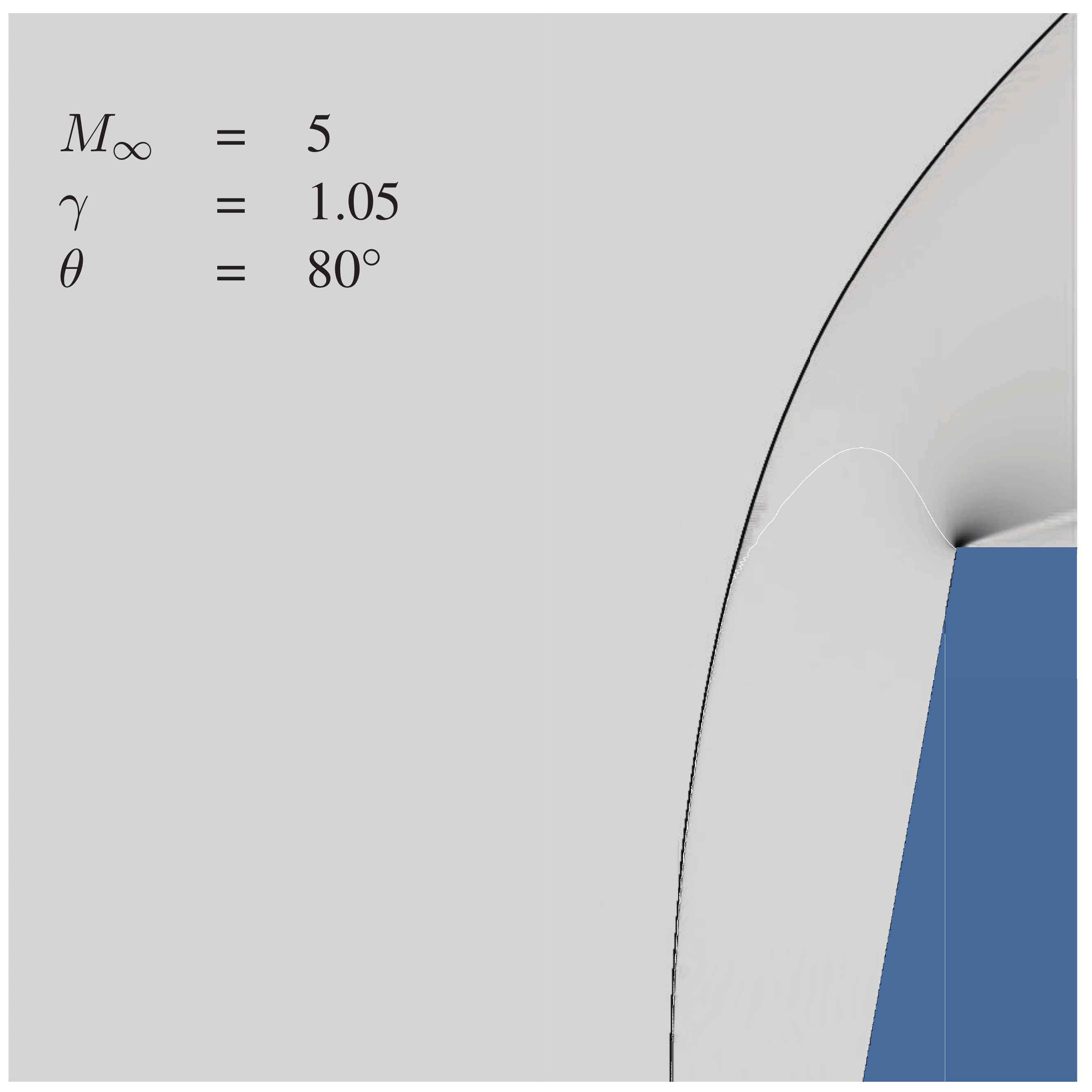} \includegraphics[trim=0 0 76 76,clip,width=0.247\columnwidth]{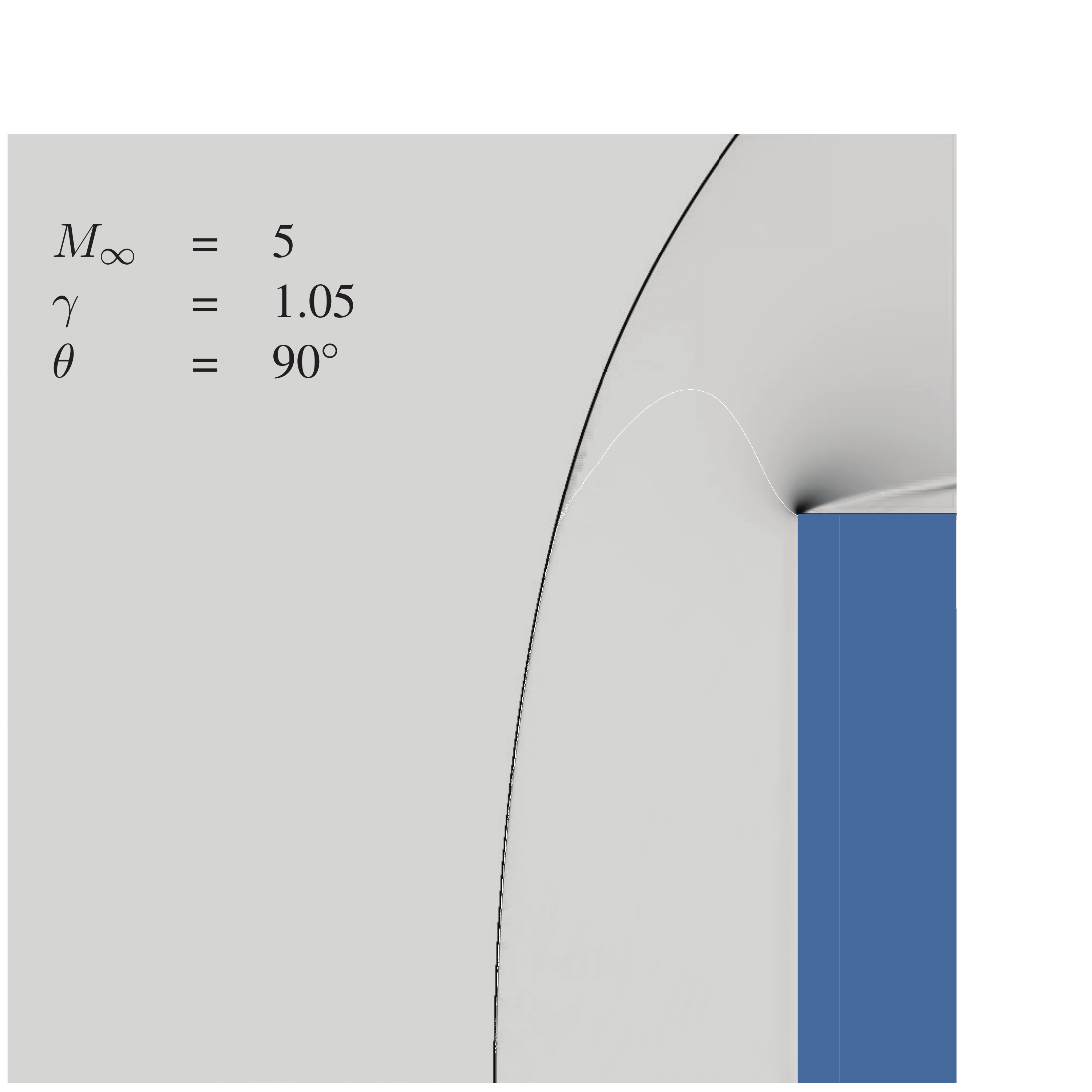}\\
\includegraphics[width=0.24\columnwidth]{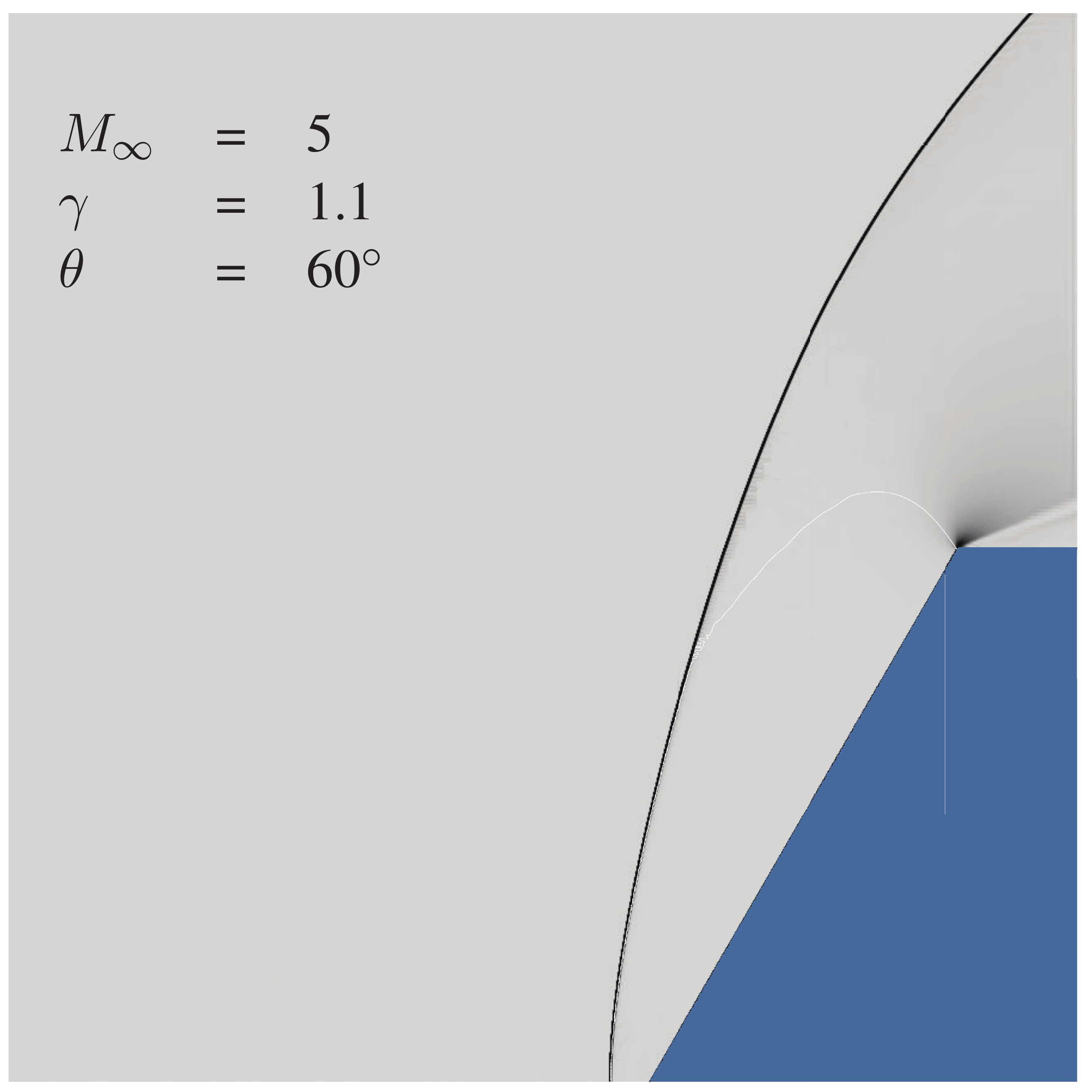} \includegraphics[width=0.24\columnwidth]{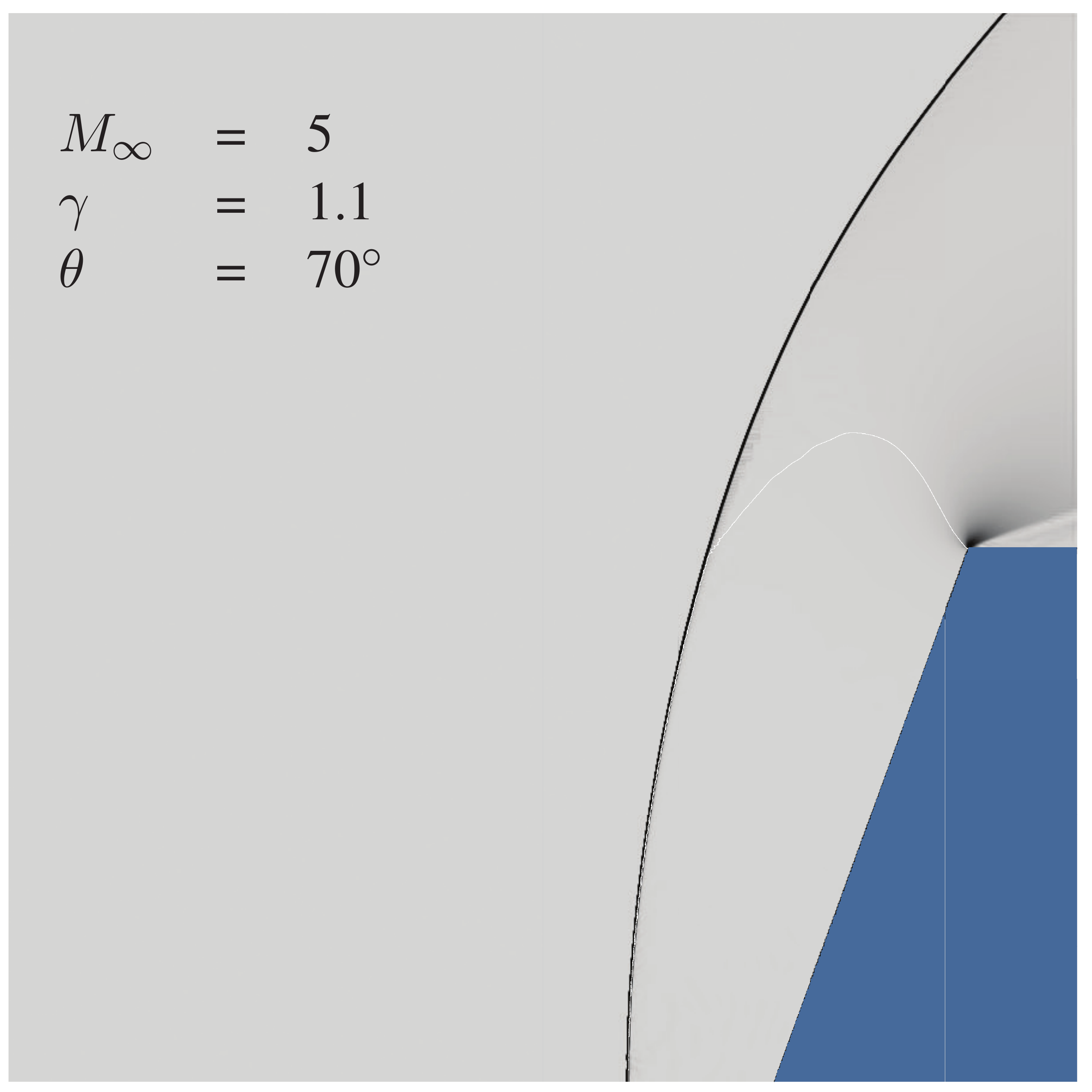} \includegraphics[width=0.24\columnwidth]{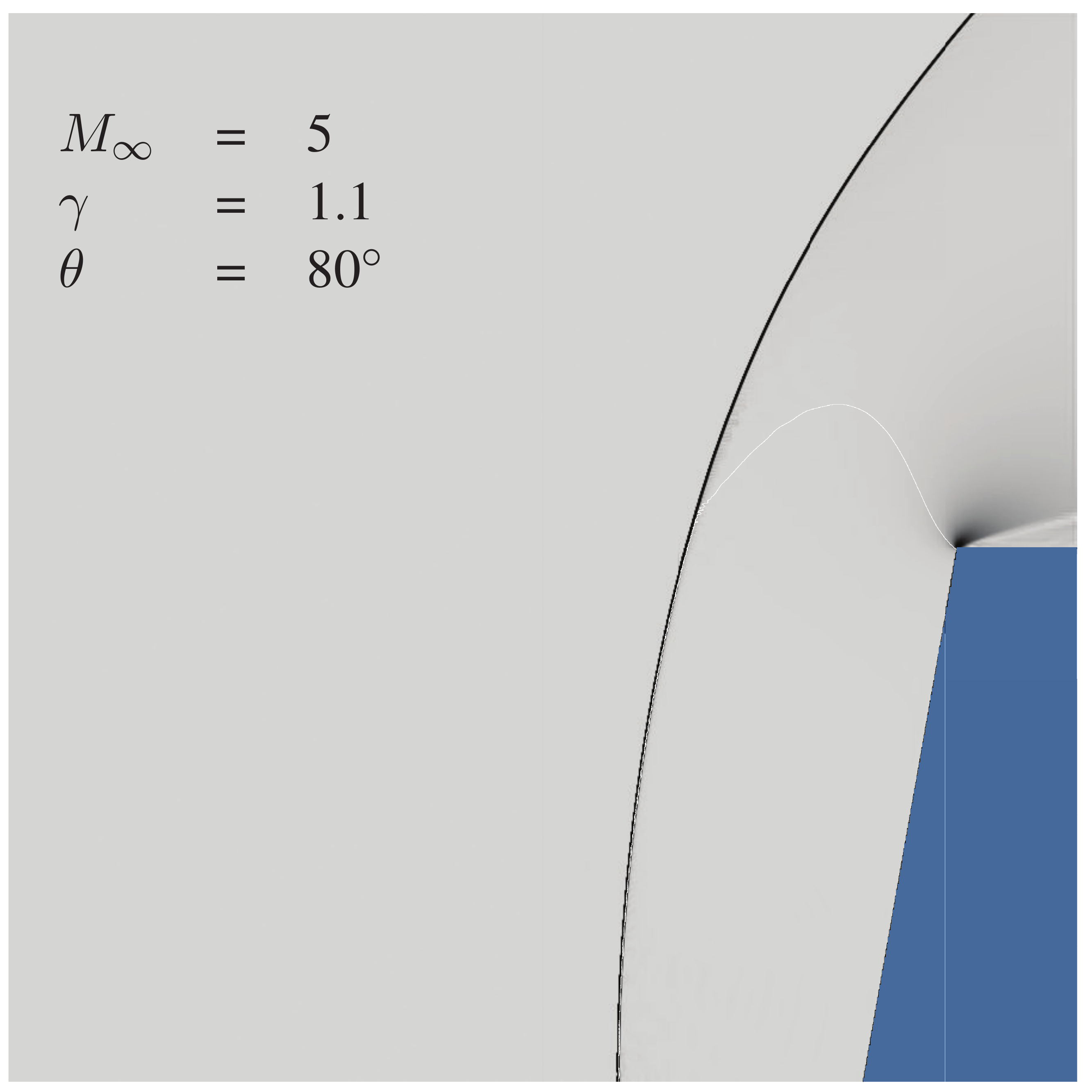} \includegraphics[trim=0 0 76 76,clip,width=0.247\columnwidth]{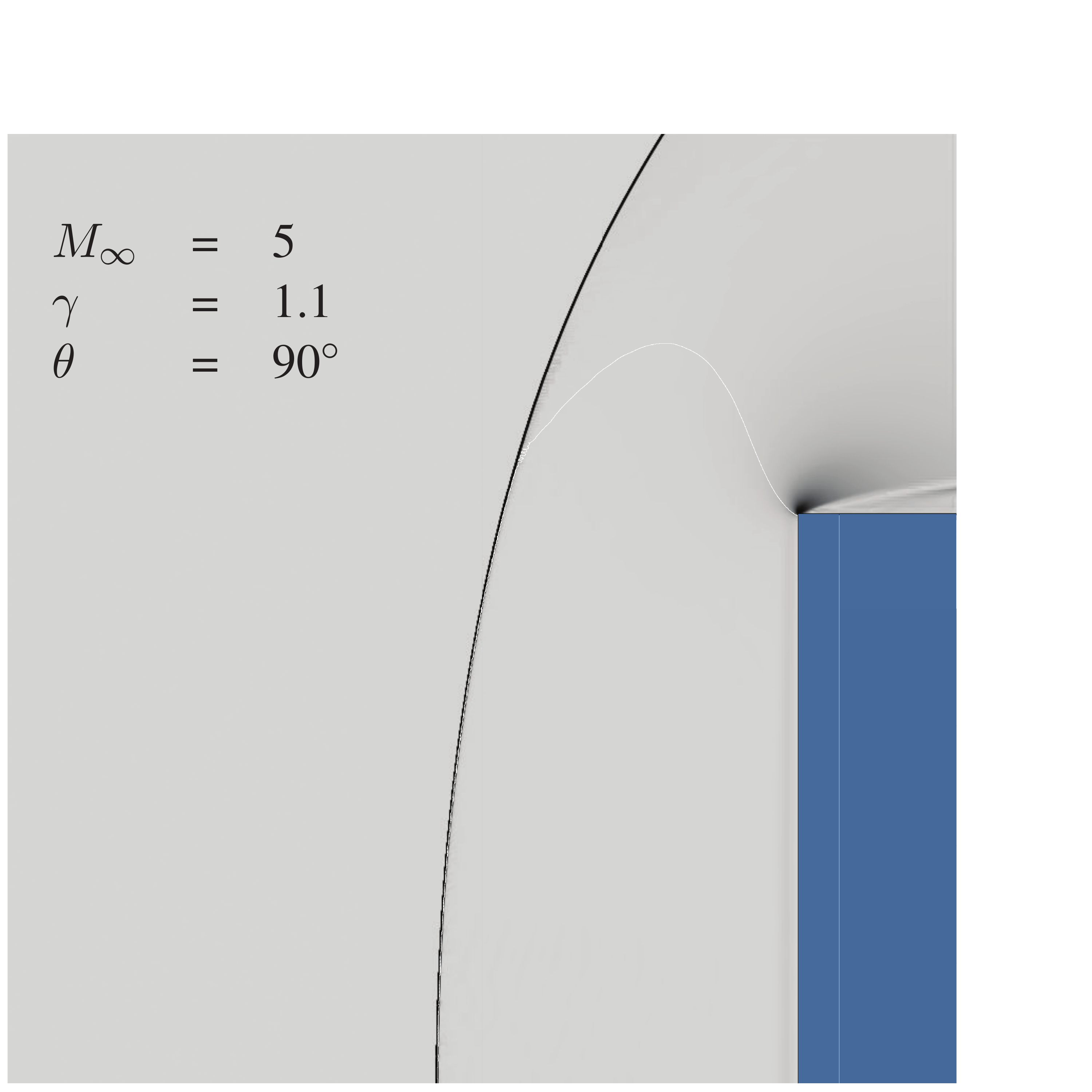}\\
\includegraphics[width=0.24\columnwidth]{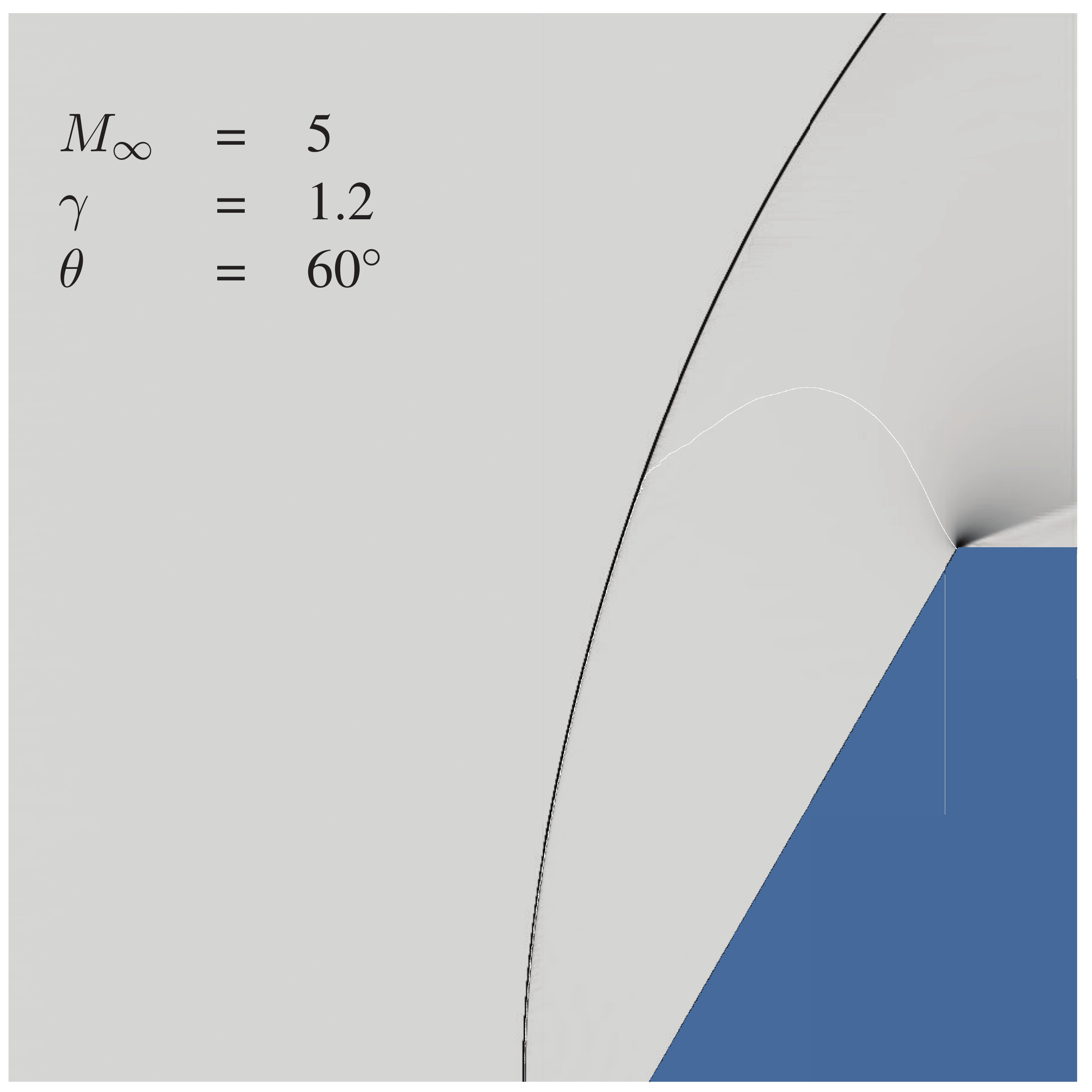} \includegraphics[width=0.24\columnwidth]{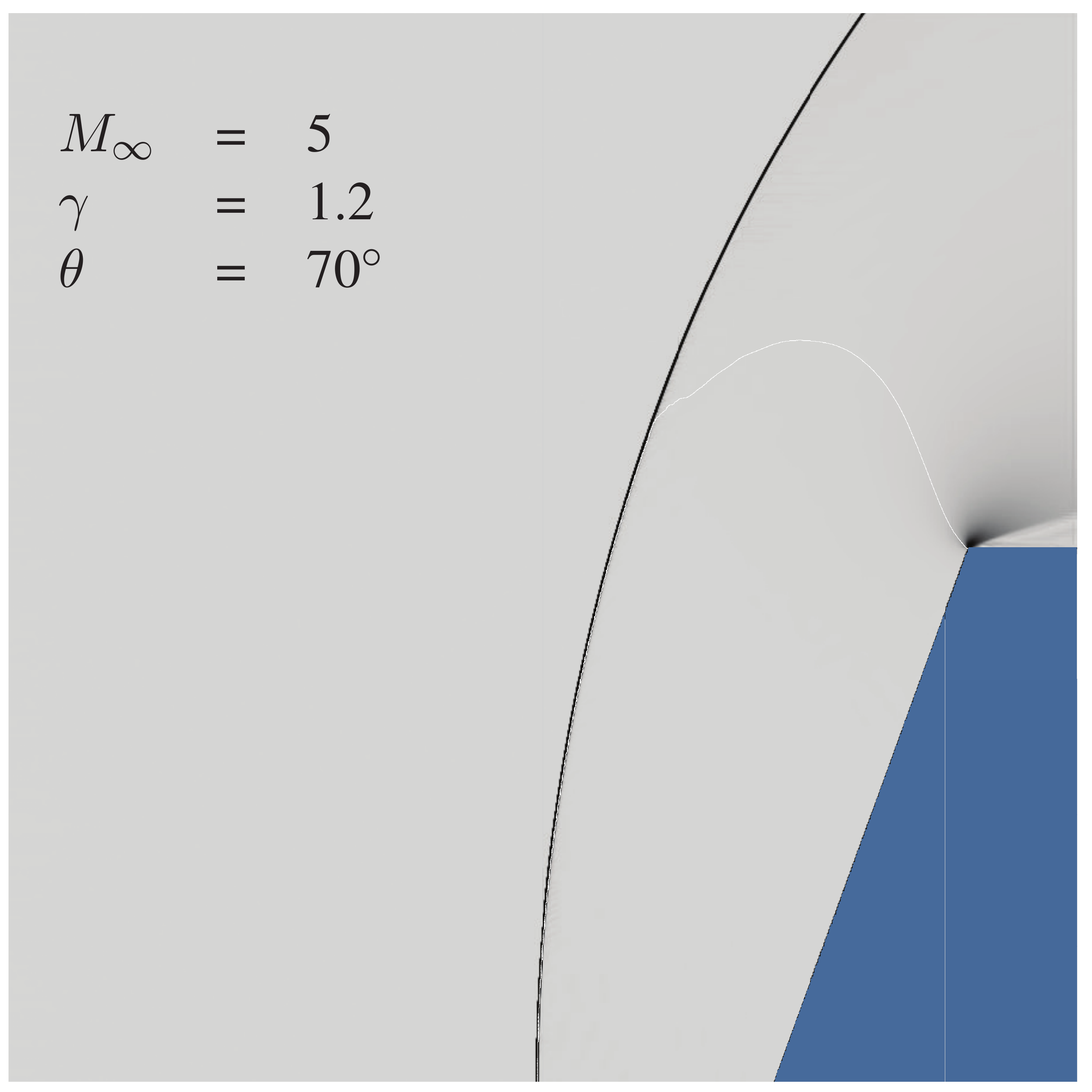} \includegraphics[width=0.24\columnwidth]{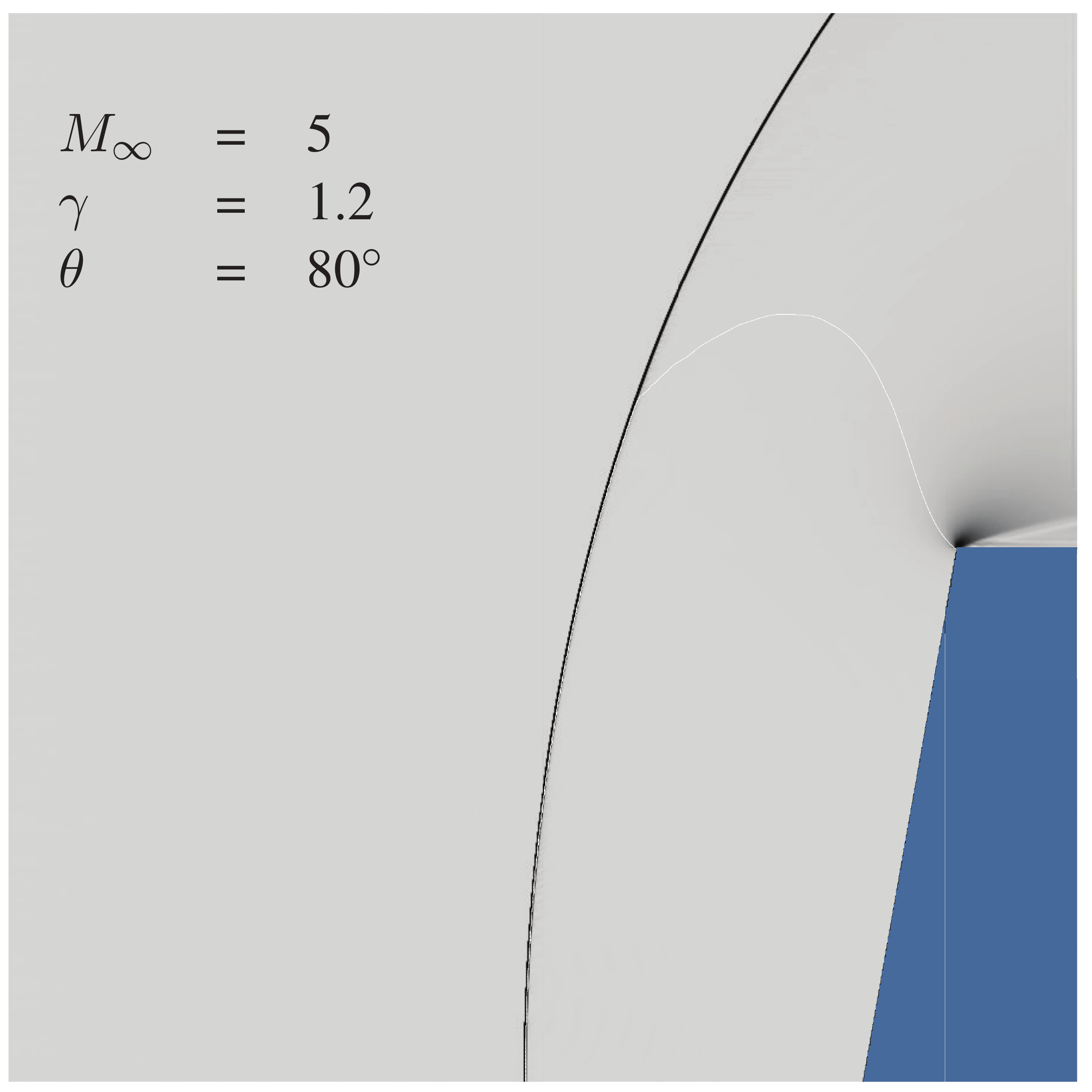} \includegraphics[trim=0 0 76 76,clip,width=0.247\columnwidth]{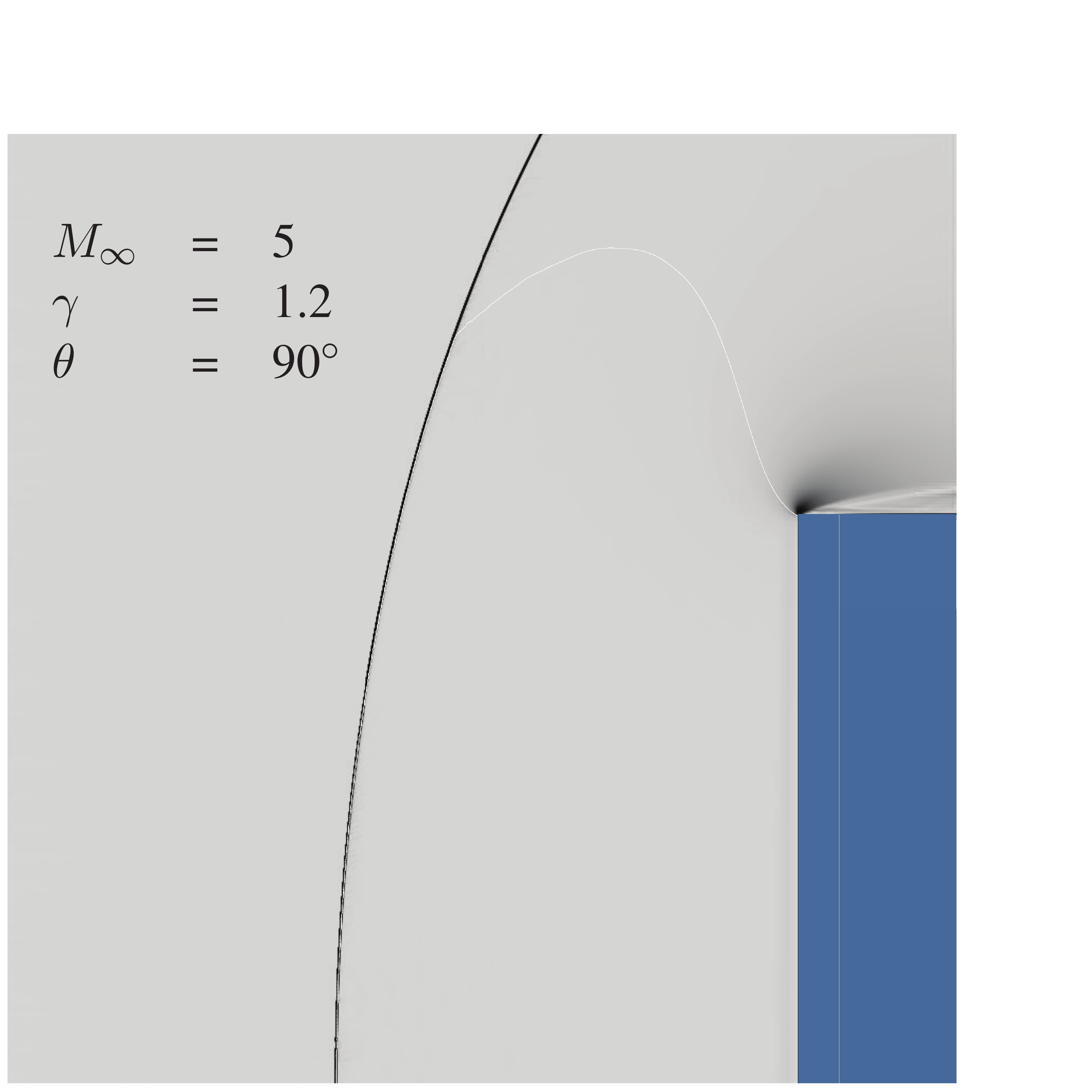}\\
\includegraphics[width=0.24\columnwidth]{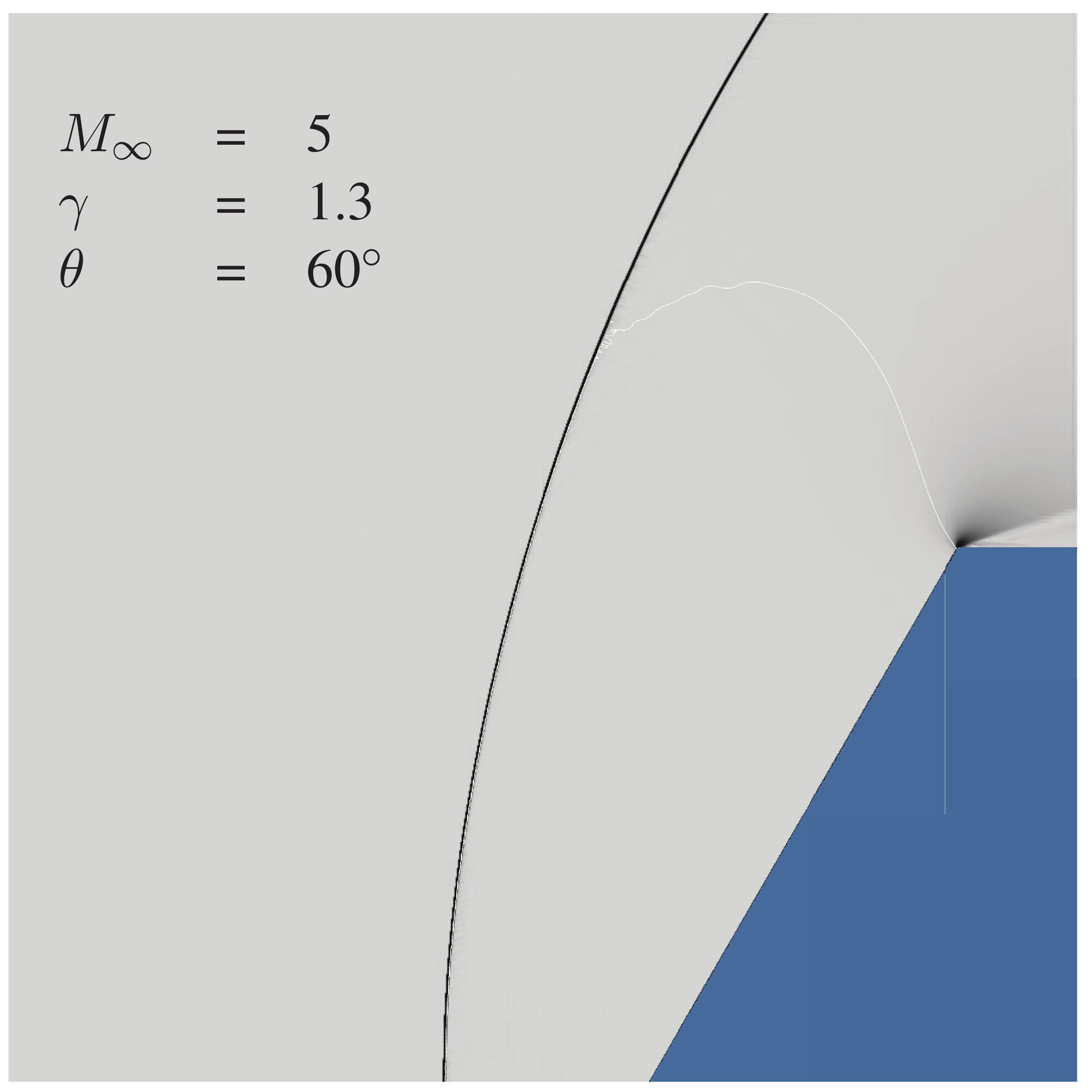} \includegraphics[width=0.24\columnwidth]{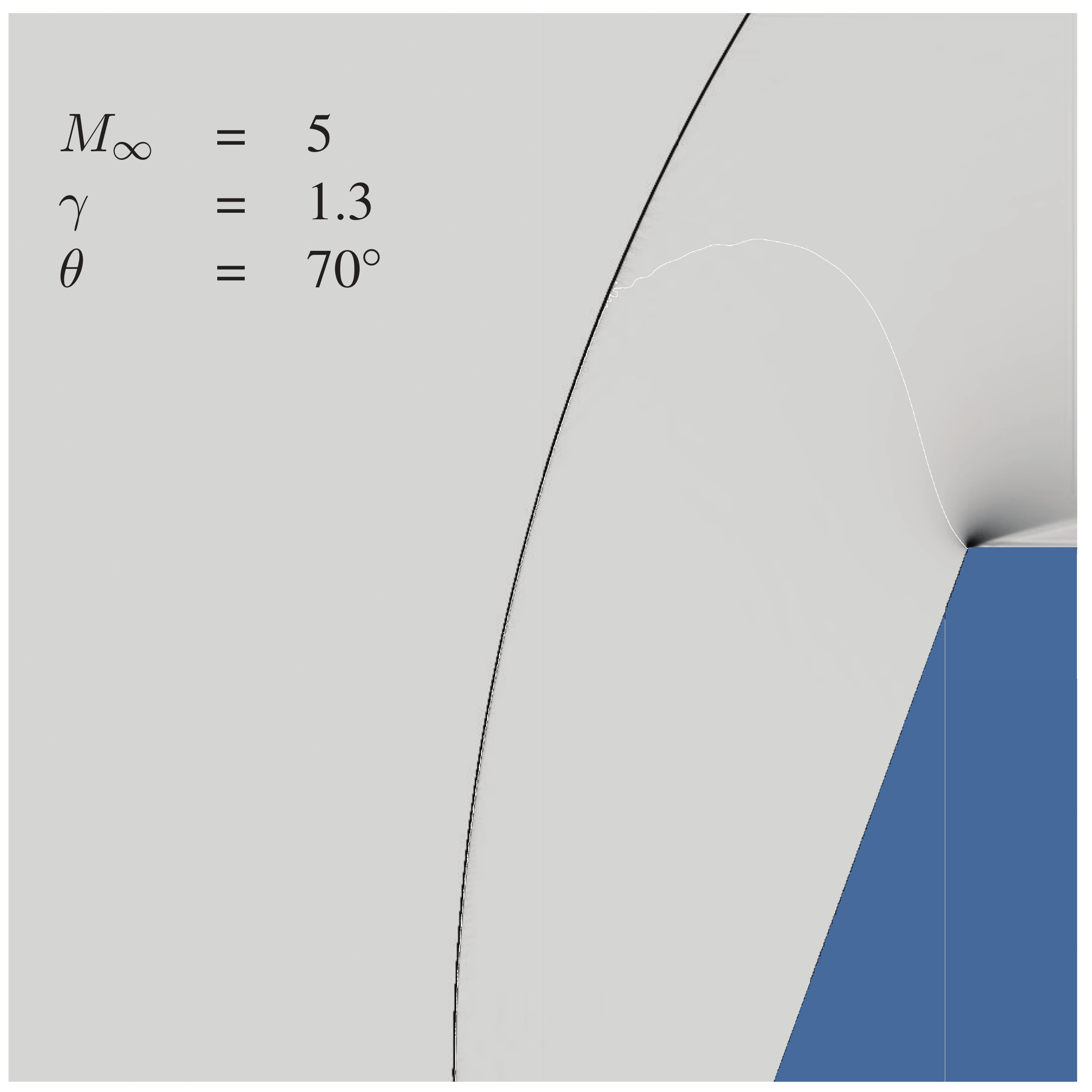} \includegraphics[width=0.24\columnwidth]{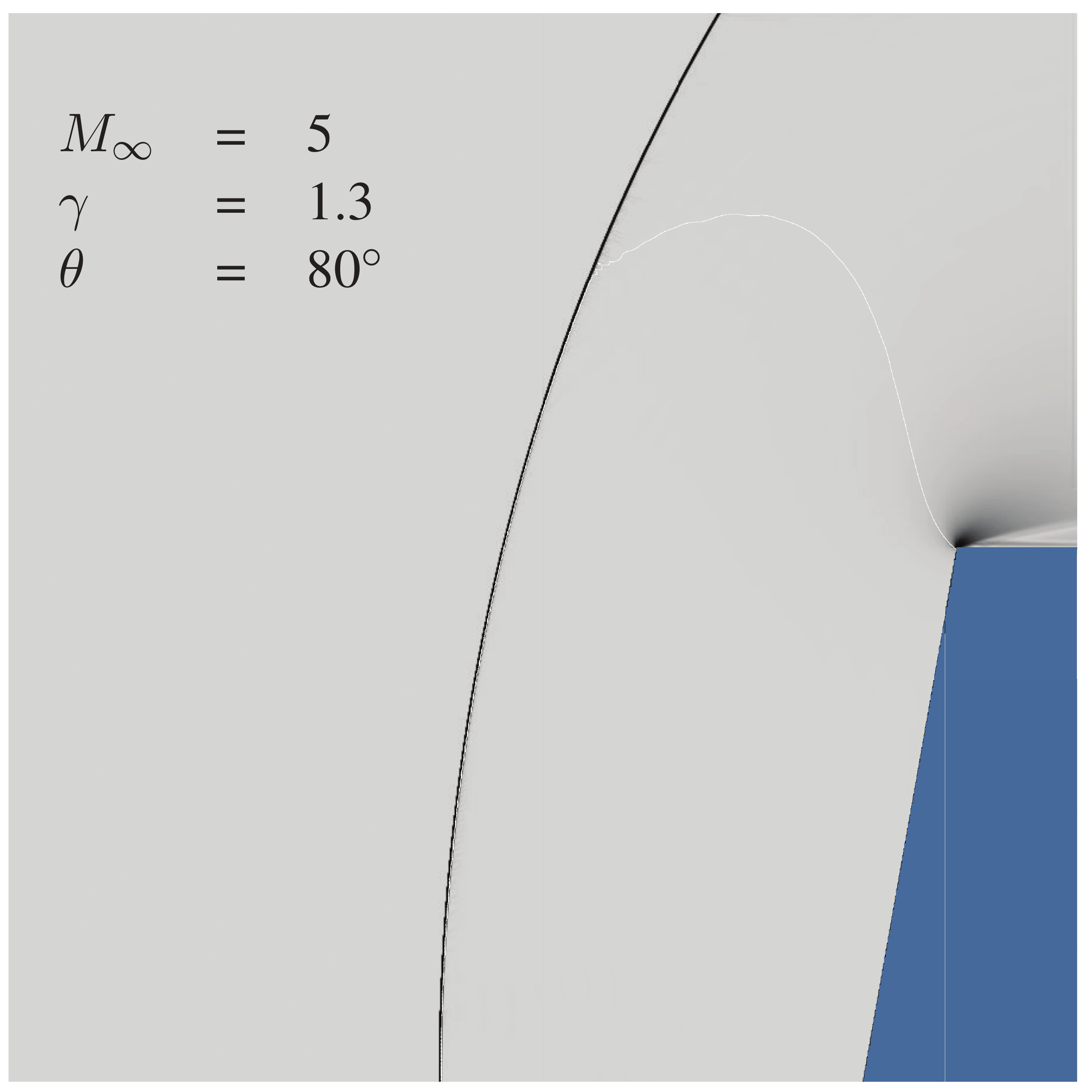} \includegraphics[trim=0 0 76 76,clip,width=0.247\columnwidth]{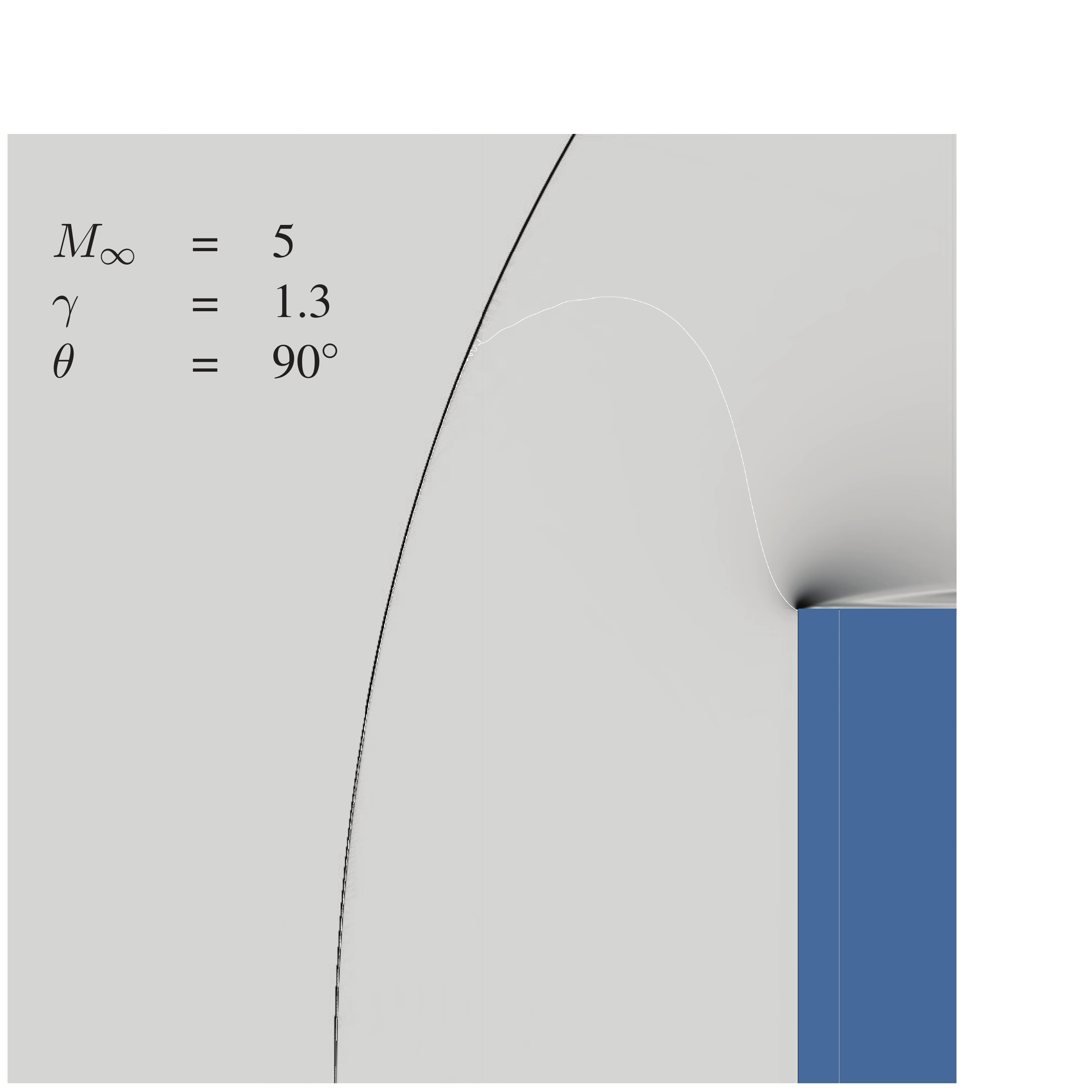}
   \end{center}
   \caption{Pseudo-schlieren images of flow over a wedge at $M_\infty=5$. 
   The grey-shading in these images is proportional to a monotonic function
   of the magnitude of the density gradient. The
   white line is the sonic line. In rows from left to right, $\theta=60$,
   70, 80 and 90 deg. In columns from top to bottom, $\gamma=1.05$, 1.1, 1.2 
   and 1.3. Similar sets of computations were made for $M_\infty=4$, 7 and 10.} 
\end{figure}
\section{Discussion of results of computations}
\subsection{Shock stand-off distance}
The parameter space was explored
by computing the flow over wedges using the Euler equations. Details about the
computational technique are given in the Appendix. An example of the results is
presented in Figure~3 for the case of $M_\infty=5$. Similar results were also
obtained for $M_\infty=4$, 7 and 10. Plotting the dimensionless shock
stand-off distance against $\varepsilon$ in the case $\theta=90^\circ$,
(square slab),
i.~e., for $\eta_e=1$, where $f(\eta_e)=f(1)$ is a constant,
provides a partial test of the hypothesis for the
function $g(\varepsilon)$.
\begin{figure}
   \begin{center}
   \includegraphics[width=0.48\columnwidth]{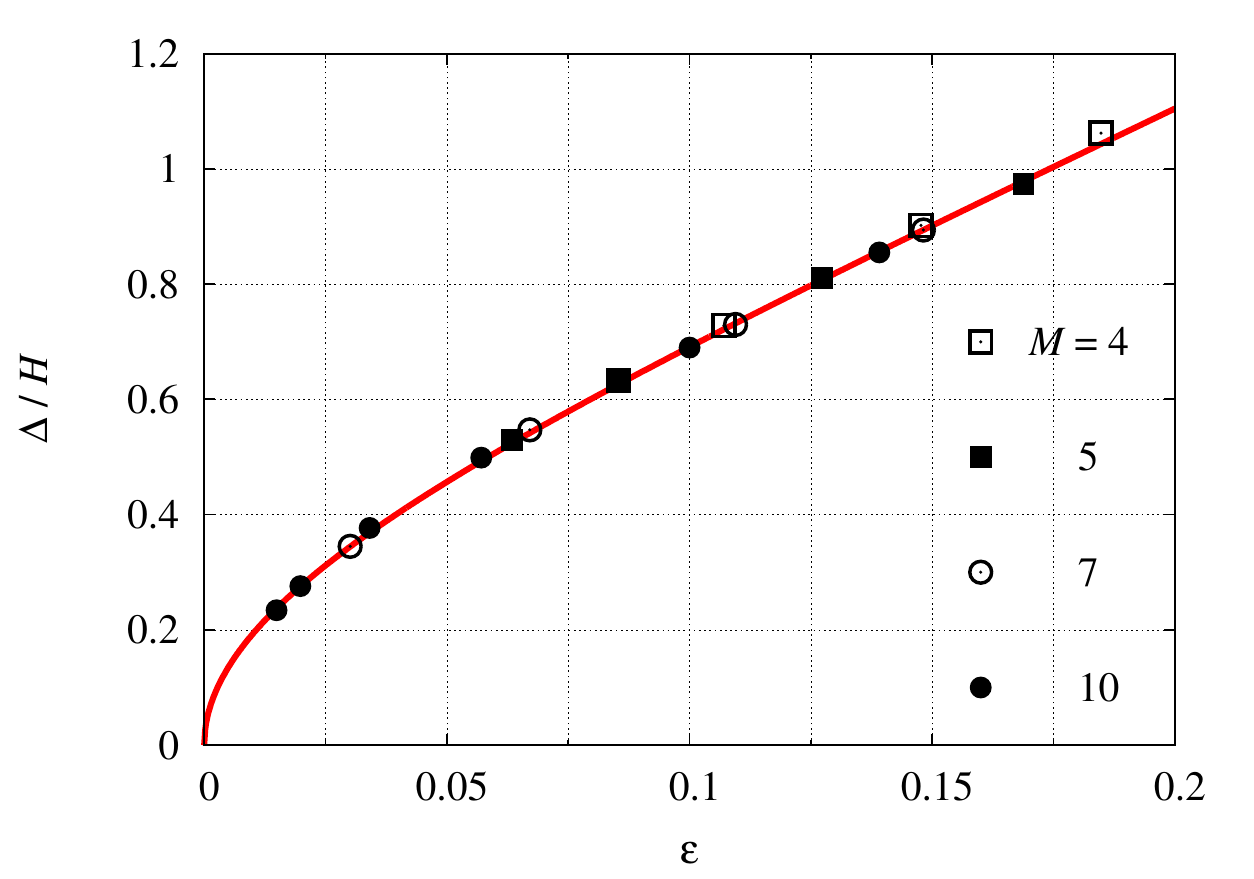} \includegraphics[width=0.48\columnwidth]{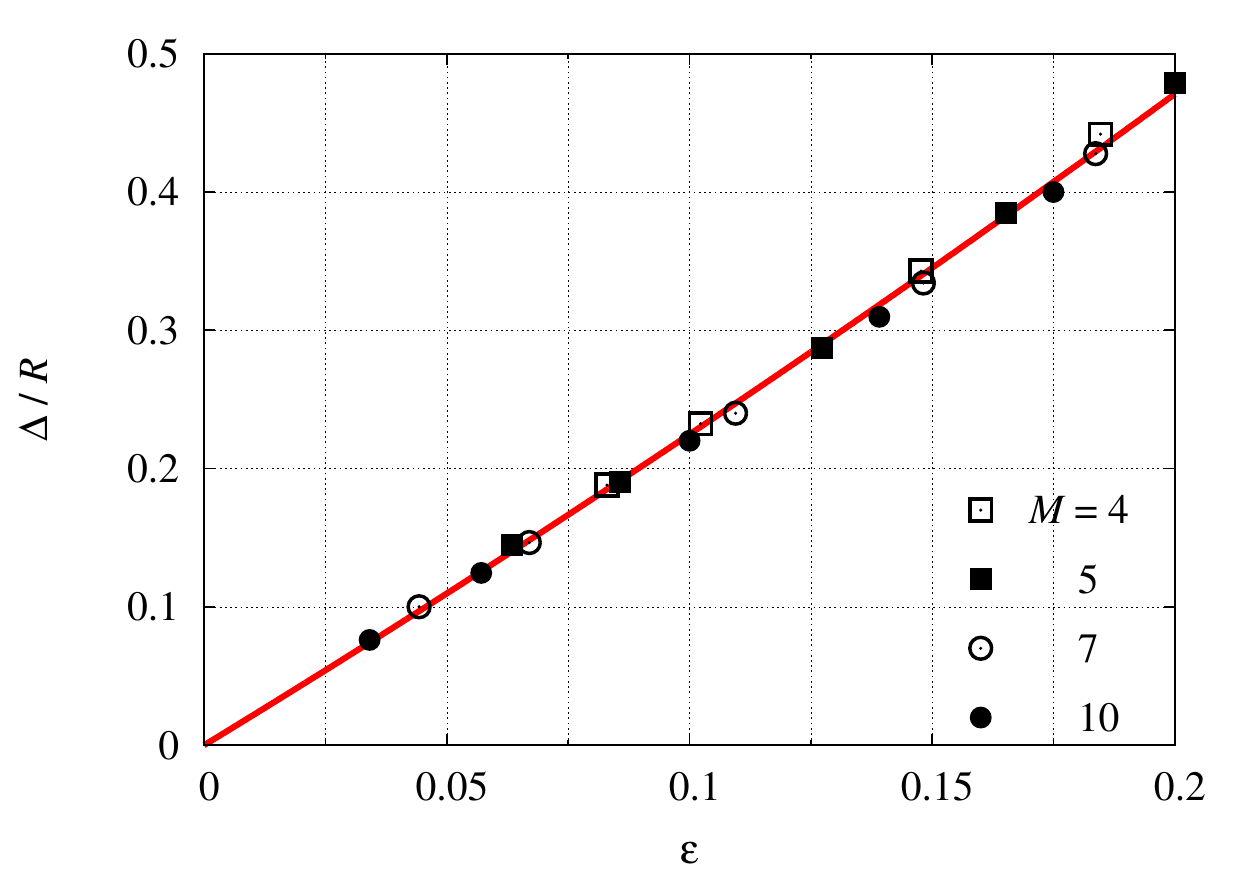}
   \end{center}
   \caption{LEFT: Dimensionless shock stand-off 
   distance for flow over a 90$^\circ$ 
   wedge with $M_\infty=4$, 5, 7 and 10 and $\gamma=1.05$, 1.1, 1.2 and 1.3.
   In the case of $M_\infty=10$, two cases of $\gamma=1.01$ and 1.02,
   and with $M_\infty=7$, one case with $\gamma=1.02$ were added.
   RIGHT: Corresponding plot for flow over a circular 
   cylinder of radius $R$.  In this case the results are fitted well by 
   $\Delta/R=2.14\,\varepsilon(1+\varepsilon/2)$. 
   In \cite{Hornung72} the linear form 
   $\Delta/R=2.32\,\varepsilon$ was found, which is only very slightly 
   different.}
\end{figure}
\begin{figure}
   \begin{center}
   \includegraphics[width=0.48\columnwidth]{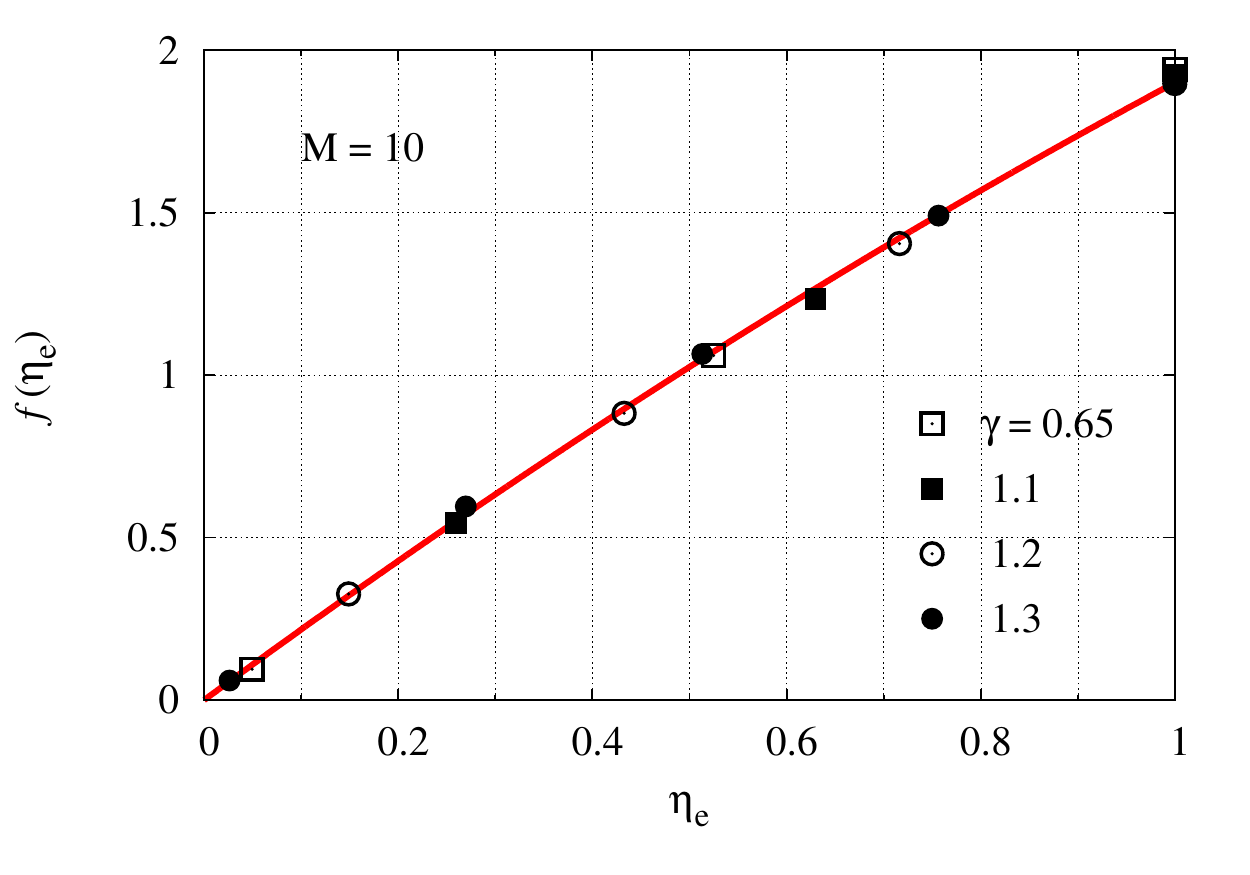} \includegraphics[width=0.48\columnwidth]{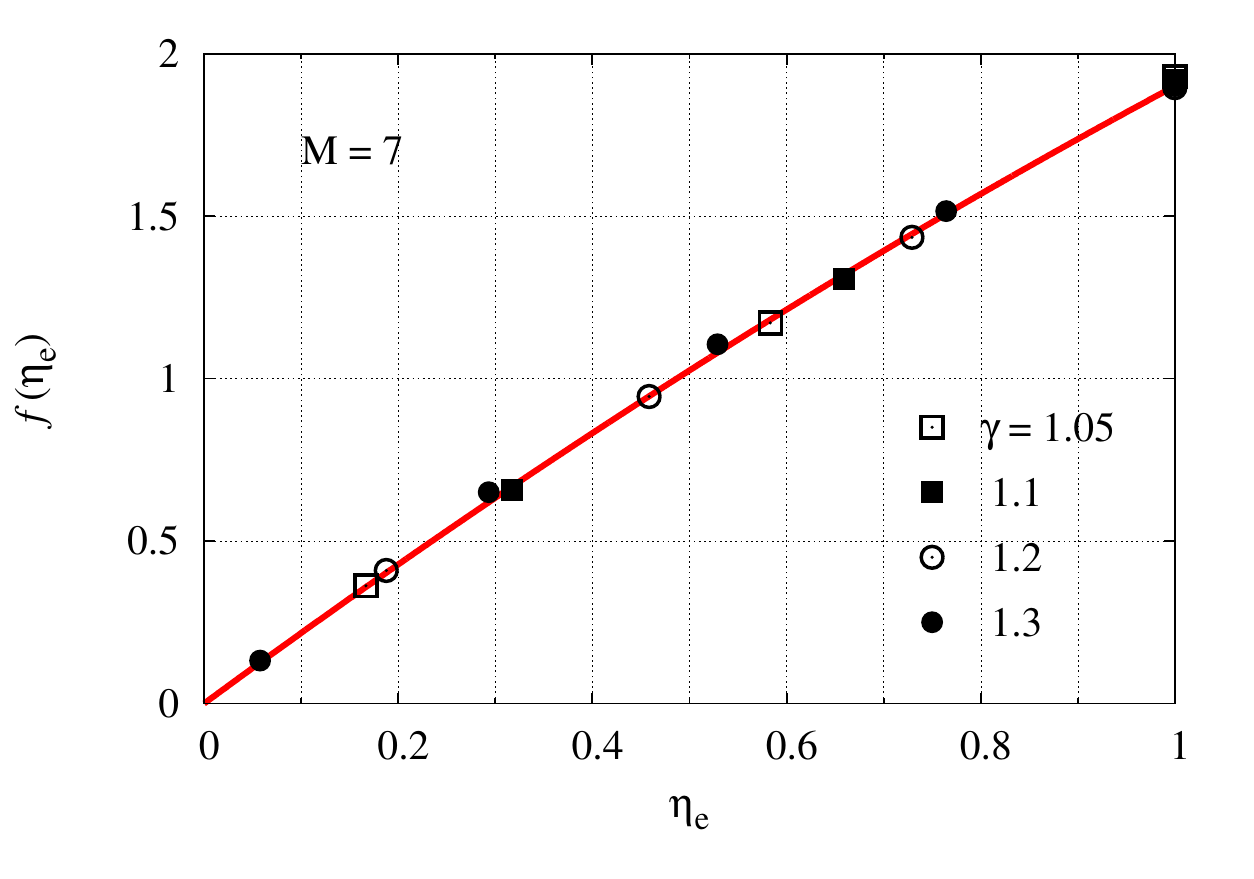}\\
   \includegraphics[width=0.48\columnwidth]{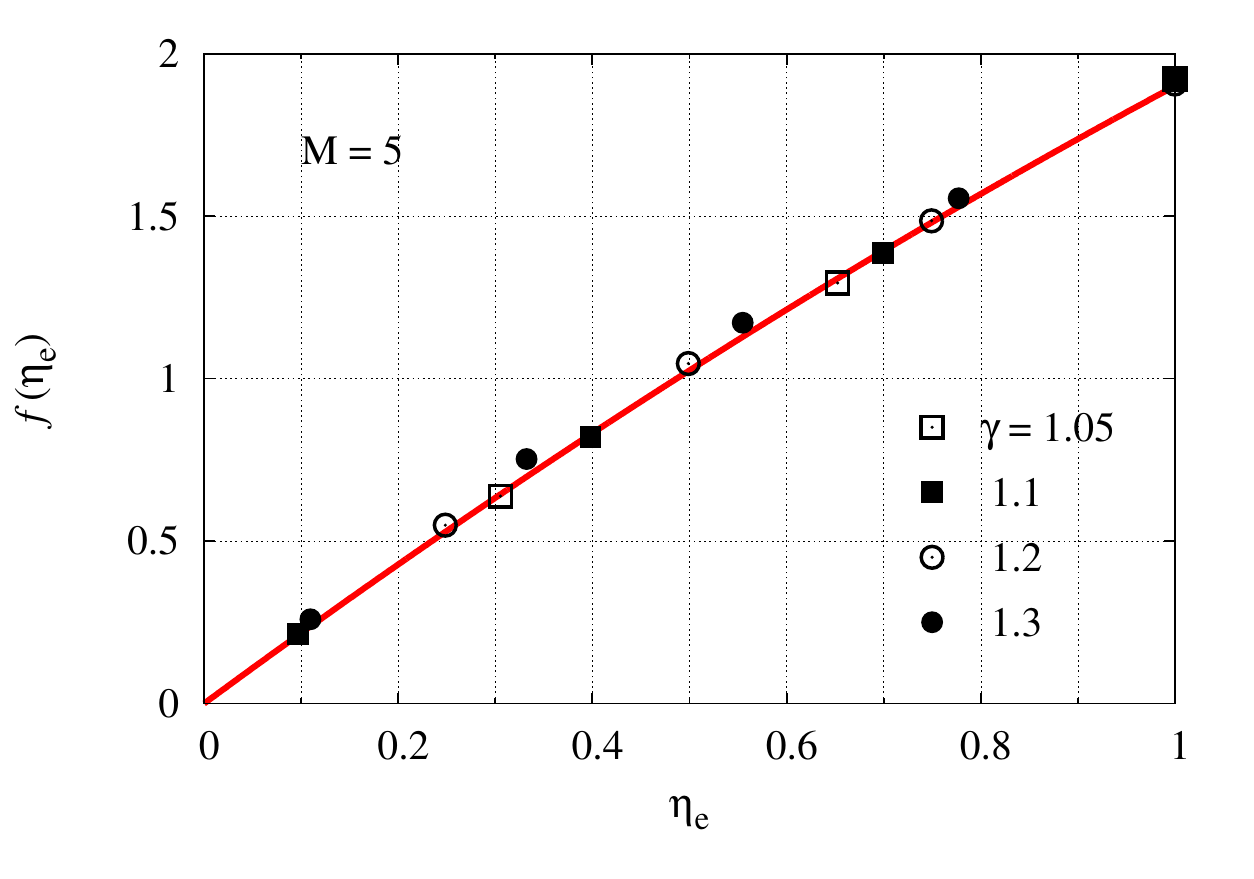} \includegraphics[width=0.48\columnwidth]{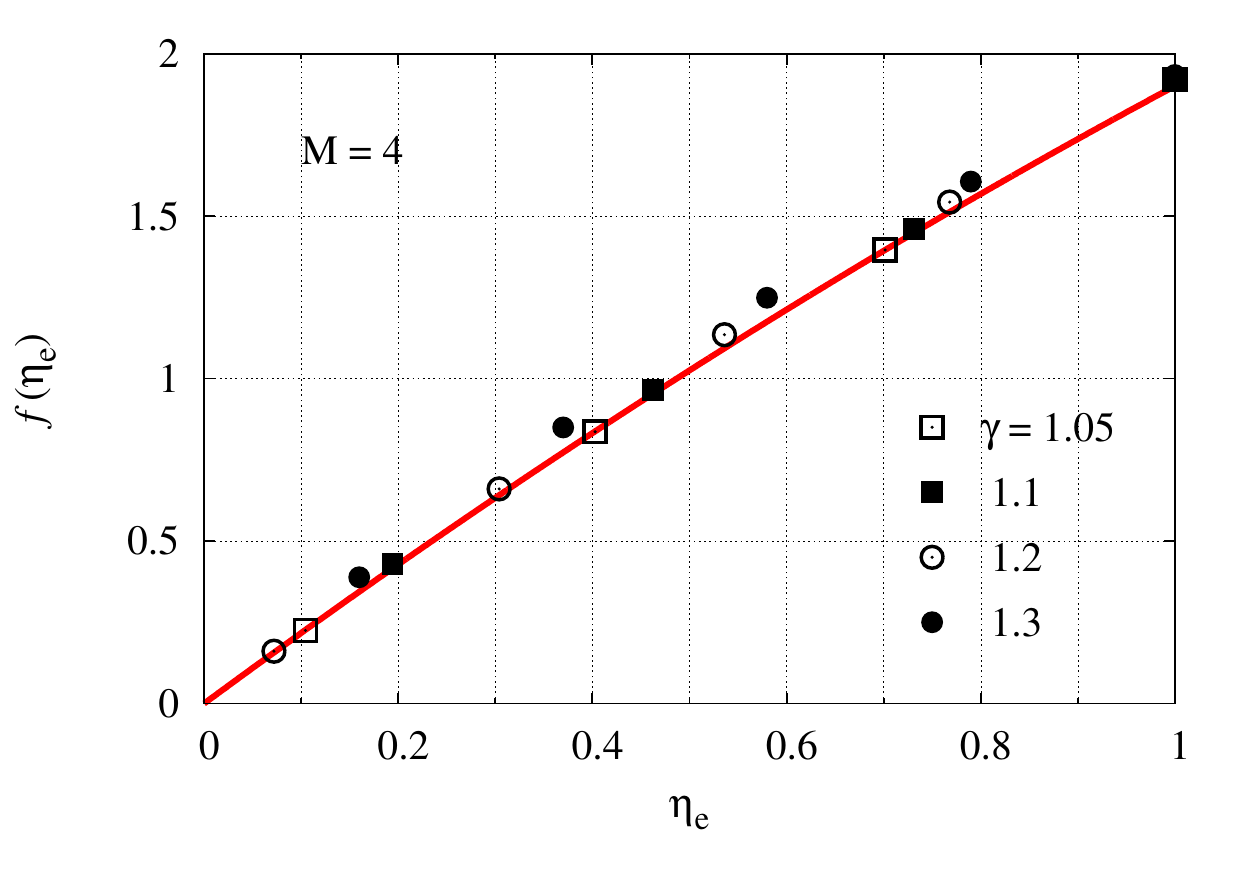}
   \end{center}
   \caption{Four plots of $f(\eta_e)$ vs. $\eta_e$. Top, left to right:
   $M_\infty=10$ and 7. Bottom, left to right $M_\infty=5$ and 4. In all
   four plots the curve is the same and given by equation~9.}
\end{figure}

This has been done in Figure~4.
A fit to the points in Figure~4 yields the interesting result that all
points fall on a unique curve given by
\begin{equation}
g(\varepsilon)\,=\,\sqrt{\varepsilon}\left(1+{3\over{2}}\varepsilon\right),
\end{equation}
thus confirming the first part of the hypothesis, i. e., that a unique function
$g(\varepsilon)$ exists.
Again, as in the case of flow over cones, the leading term is 
proportional to $\sqrt{\varepsilon}$ and the fact that all the results
fall on the same curve confirms the first part of the hypothesis. It is
interesting that in the corresponding function of $\varepsilon$ for 
the 90$^\circ$ cone, the factor 3/2 that appears in equation~8 is 1/2. 

To test the 
second part, four plots of $f(\eta_e)=\Delta/[H\,g(\varepsilon)]$ vs.
$\eta_e$ are shown in Figure~5 for the four Mach numbers and the four
$\gamma$'s. All the computational results fall on the same curve given by the
 unique function
\begin{equation}
f(\eta_e)\,=\,2.2\,\eta_e-0.3\,\eta_e^2,
\end{equation}
thus confirming the second part of the hypothesis. Note however, that the
fit is not so good at $M_\infty=4$ as for the others. Had we used $\eta$ instead
of $\eta_e$, this deterioration of the fit at the lower Mach numbers would have
been slightly larger.

\subsection{Drag coefficient}
In \cite{Hornung2019} it was shown that for flow over cones, the drag 
coefficient could also be expressed in the form of equation~5. Write the drag
coefficient for a wedge as
\begin{equation}
C_D\,=\,{2D\over{\gamma p_\infty M_\infty^2 H L}},
\end{equation} 
where the drag force
$$D\,=\,2L\int_0^H(p-p_\infty)dy,$$
$L$ is the transverse length of the wedge, and $y$ is the distance measured
from the symmetry plane of the wedge.
Then, if the form of equation~5 holds for $C_D$, expect that
\begin{equation}
C_D\,=\,g_1(\varepsilon) f_1(\eta_e).
\end{equation}
The same set of computational results can now be used to check whether this is 
correct. Again we use the case $\eta_e=1$ where $f_1(1)$ is a constant to check
if $g_1(\varepsilon)$ is unique. To this end, Figure~6 shows a plot of $C_D$
vs. $\varepsilon$ for the 90$^\circ$ wedge and for the circular cylinder. In
both cases all the results collapse onto a single line. for the 90$^\circ$ 
wedge,
\begin{equation}
C_d\,=\,g_1(\varepsilon)\,=\,2-1.4\,\varepsilon,
\end{equation}
and for the circular cylinder
\begin{equation}
C_D\,=\,1.3-5(\varepsilon-0.085)^2.
\end{equation}
\begin{figure}[ht!]
   \begin{center}
   \includegraphics[width=0.46\columnwidth]{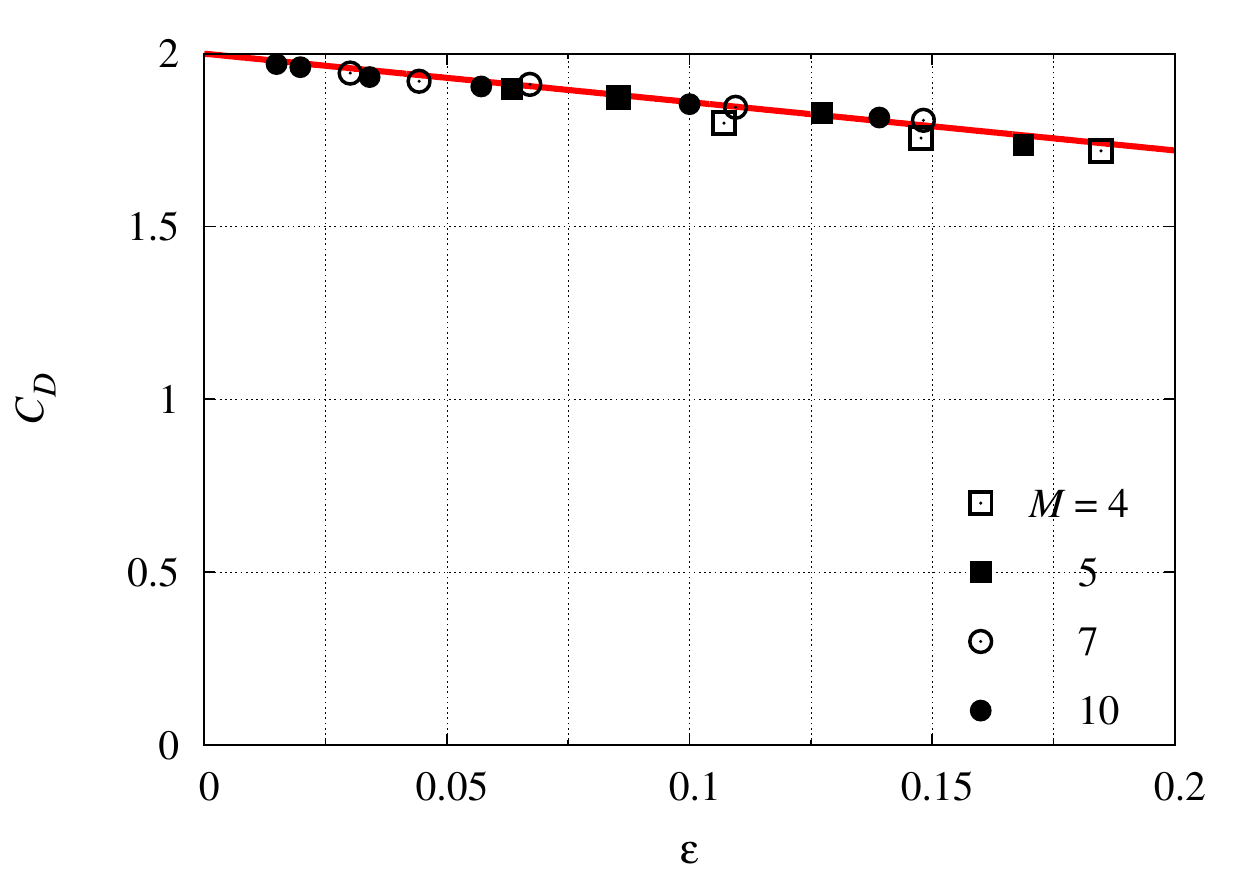} \includegraphics[width=0.46\columnwidth]{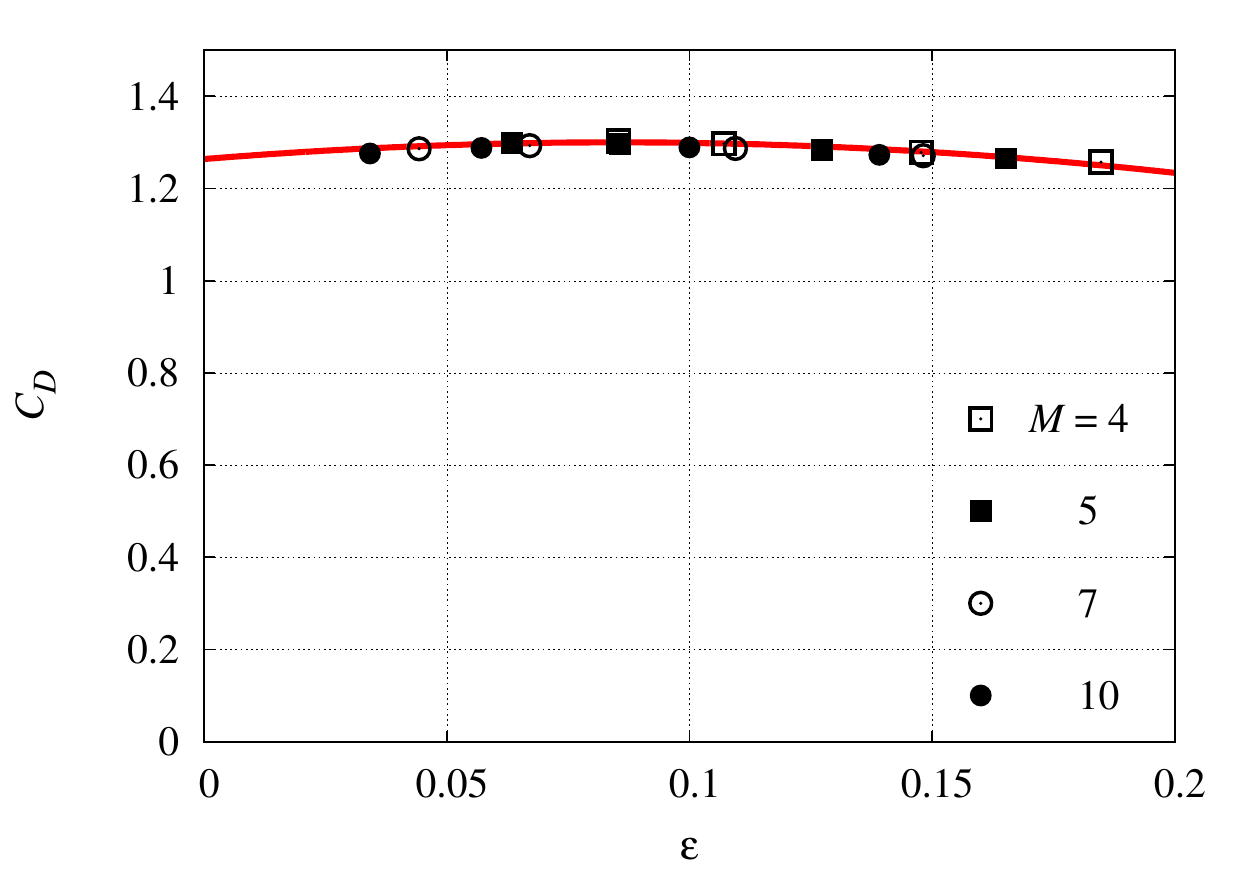}
   \end{center}
   \caption{LEFT: Drag coefficient for flow over a 90$^\circ$
   wedge with $M_\infty=4$, 5, 7 and 10 and $\gamma=1.05$, 1.1, 1.2 and 1.3.
   For $M_\infty=10$, two cases of $\gamma=1.01$ and 1.02,
   and with $M_\infty=7$, one case with $\gamma=1.02$ were added. RIGHT: Drag
   coefficient for flow over a circular cylinder with $M_\infty=4$, 5, 7 and 10 and $\gamma=1.05$, 1.1, 1.2 and 1.3.}
\end{figure}
\begin{figure}[ht!]
   \begin{center}
   \includegraphics[width=0.46\columnwidth]{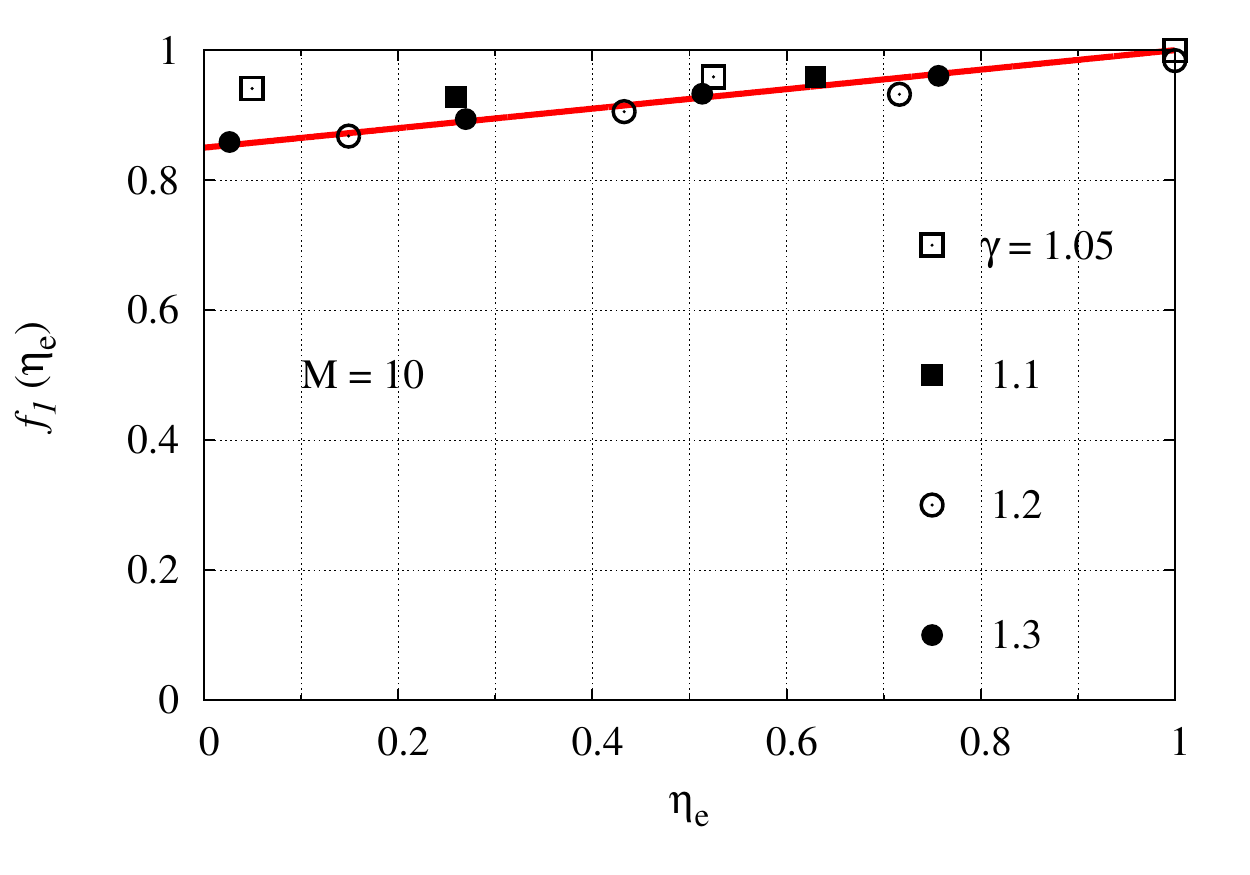} \includegraphics[width=0.46\columnwidth]{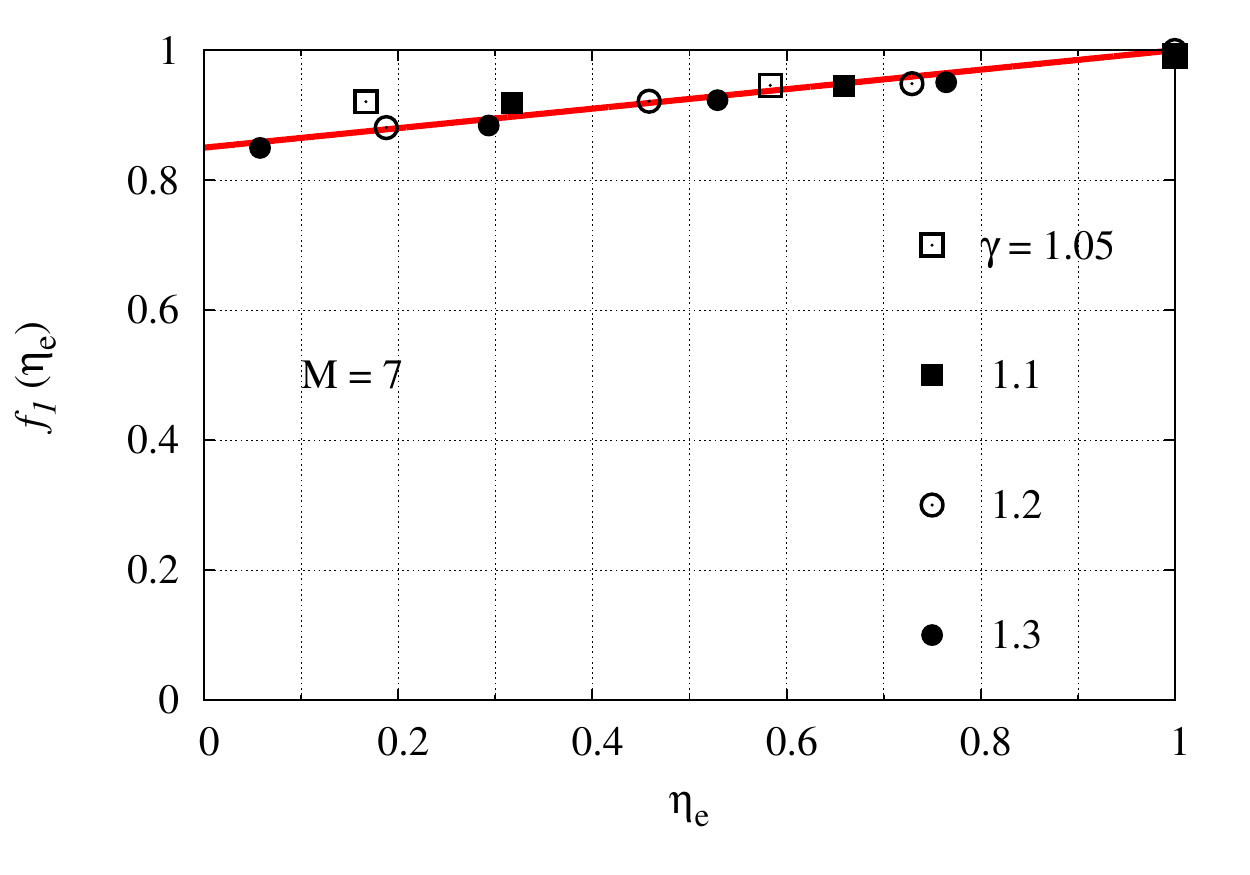}\\
   \includegraphics[width=0.46\columnwidth]{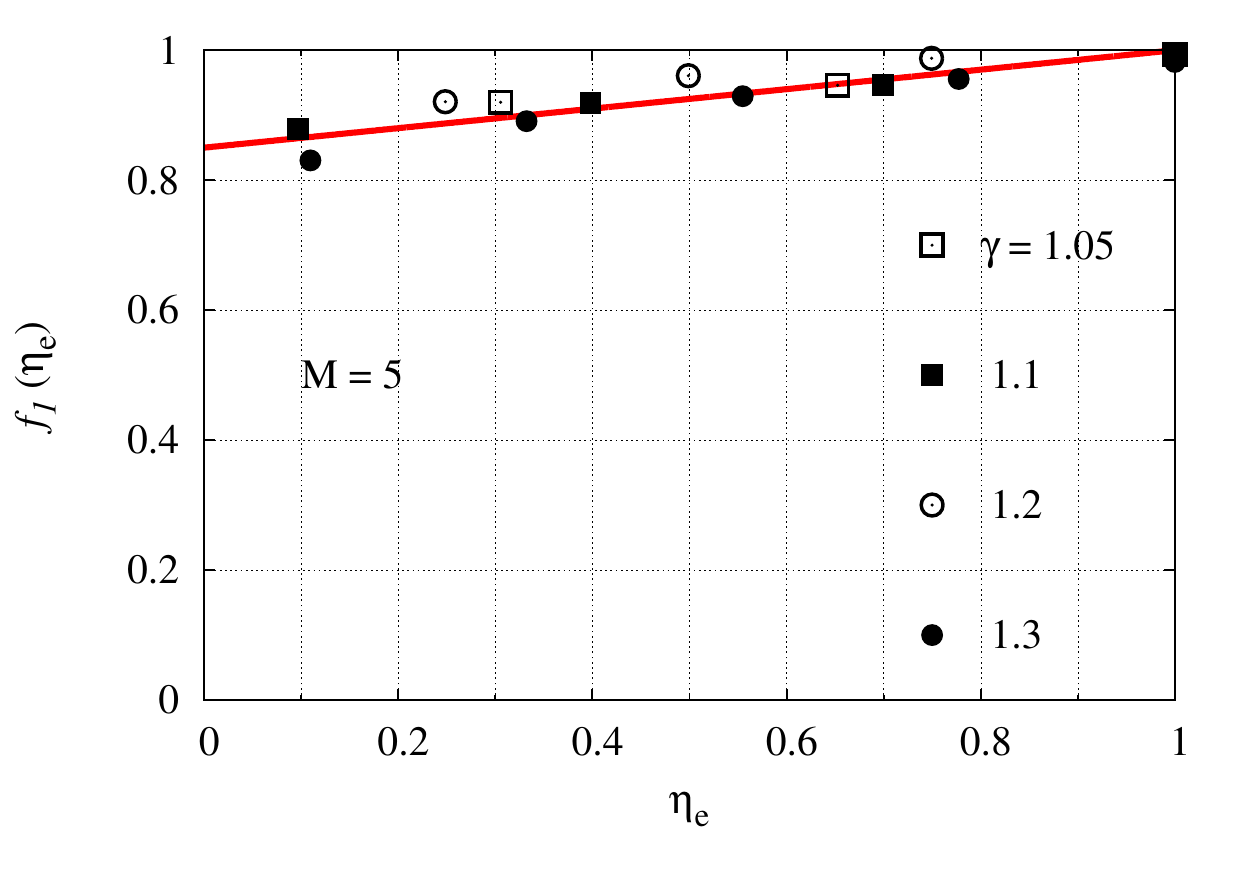} \includegraphics[width=0.46\columnwidth]{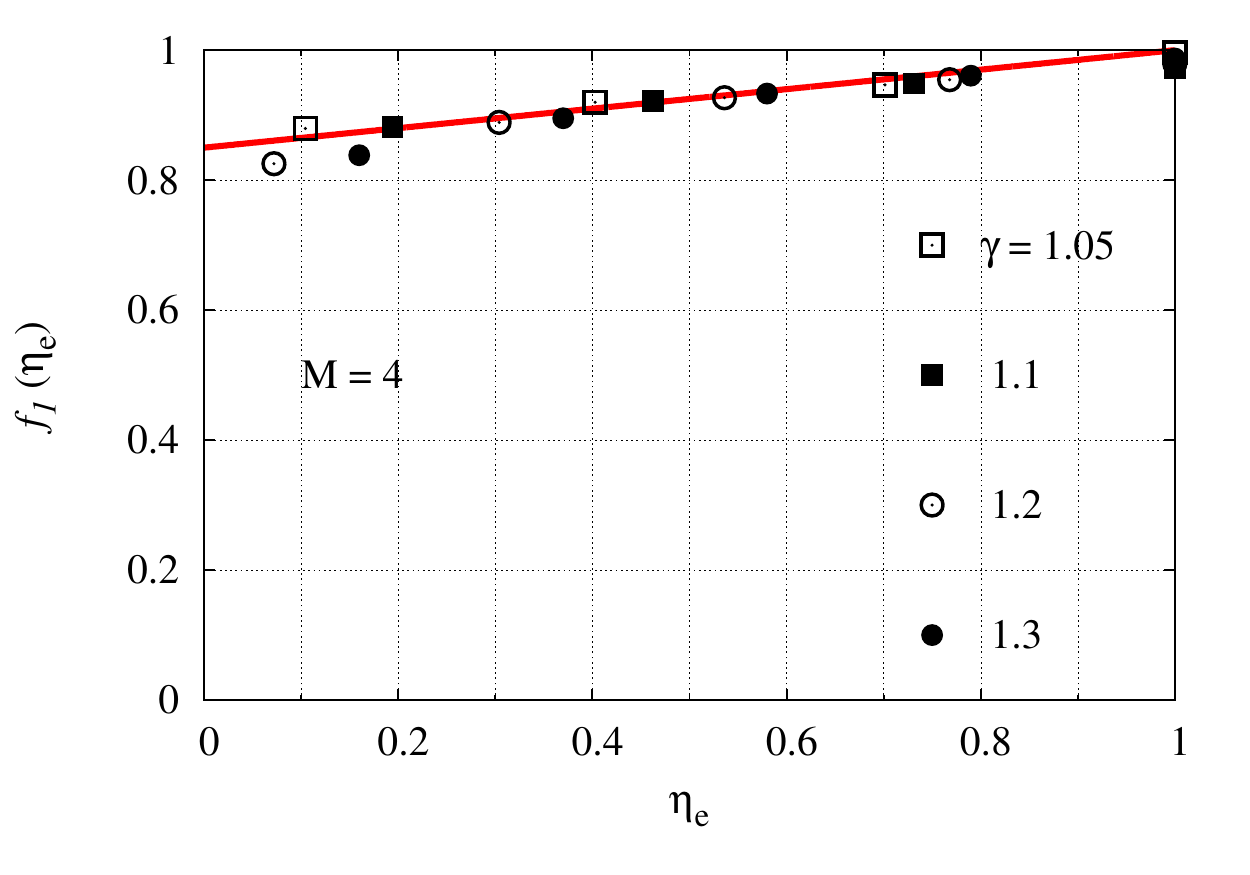}
   \end{center}
   \caption{Four plots of $f_1(\eta_e)$ vs. $\eta_e$. Top, left to right:
   $M_\infty=10$ and 7. Bottom, left to right $M_\infty=5$ and 4. In all
   four plots the line is the same and given by equation~14.}
\end{figure}
In order to determine the form of the function $f_1(\eta_e)$ Figure~7 shows four
plots of $\Delta/[Hg_1(\varepsilon)]$ vs.$\eta_e$. While the results agree
approximately with 
\begin{equation}
0.85+0.15\,\eta_e,
\end{equation}
they scatter fairly broadly around the line, so that the validity of the
functional form is not as convincing as in the case of the shock stand-off 
distance.

\section{High-enthalpy effects}
Although the results were obtained from perfect-gas computations, they may be
expected to apply also to flows at high enthalpy, where vibrational excitation 
and dissociation may occur and non-equilibrium effects become important. 
As has been shown by \cite{Wen95} and by \cite{Stulov69}, forming the 
density ratio with the average density along the stagnation streamline
instead of with the post-normal-shock density causes the results to carry over 
to the high-enthalpy regime.

\section{Conclusions}
It was shown that the reduction from three independent parameters to two,
in the parameter space defining the dimensionless shock stand-off distance and
drag coefficient, that was previously found for axisymmetric hypersonic flow 
over cones, applies also to flow over wedges in the detached-shock range.
Useful analytical forms were found for these relations by performing a large 
number of Euler computations and plotting the results in the appropriate form.
The results are also compared with new results for flow over a circular 
cylinder.
\vskip 0.2in

\noindent{\bf\large Acknowledgement}\\[10pt]
This work was funded by AFOSR FA9550-19-1-0219, PI:  J. M. Austin, Contract
Officer:  I. A. Leyva.
\vskip 0.2in

\noindent{\bf\large Appendix}\\[10pt]
The software system Amrita,
constructed by James Quirk, see \cite{Quirk98},
was used. A detailed description of the features
and phenomena encountered with some of the algorithms used for
Riemann solvers, including the one used here, has also been given by
\cite{Quirk94}. An example of a test of the software against experiment
may be found in \cite{QuirkKarni}.

Amrita is a system that automates and packages
computational tasks in such a way that the packages
can be combined (dynamically linked) according to
instructions written in a high-level scripting language.
The present application uses features of Amrita
that include the automatic construction of the Euler
solver, documentation of the code,
adaptive mesh refinement according to
simply chosen criteria, and scripting-language-driven
computation, archiving and post-processing of the results.
The automation
of the assembly and sequencing of the tasks makes
for dramatically reduced possibility of hidden errors.
It also makes computational investigations
transparent and testable by others. The ability to change
one package at a time, without changing the rest of the
scheme,  facilitates detection of sources of error.
In
most of the work, the Euler solver generated was an operator-split scheme with HLLE
flux (after Harten et al.\cite{Harten83} and Einfeldt \cite{Einfeldt88})  and kappa-MUSCL
reconstruction.
In some cases with $\gamma$ close to 1, the carbuncle problem
arose, and the more robust equilibrium flux method of \cite{Pullin80}
was used. 
The ($x,y$) plane was discretized by a Cartesian grid of
300$\times$300 coarse--grid cells that are adaptively refined by a factor of 3
to make an effective grid of 900$\times$900 cells. The criterion for
adaptation was a chosen threshold of the magnitude of the
density gradient. Solid boundaries are represented by a level set
defined as the smallest distance of a field point from the solid boundary.
The grey-shading of the visualizations is a monotonic function of the
magnitude of the density gradient.

\bibliographystyle{hplain}
\bibliography{wedge_detach}

\begin{thebibliography}{10}

\bibitem{Einfeldt88}
B.~Einfeldt.
\newblock On {G}odunov-type methods for gas dynamics.
\newblock {\em SIAM J. Numer. Anal.}, 25:294--328, 1988.

\bibitem{Harten83}
A.~Harten, P.~D. Lax, and B.~van Leer.
\newblock On upstream differencing and {G}odunov-type schemes for hyperbolic
  conservation laws.
\newblock {\em SIAM Rev.}, 25:35--61, 1983.

\bibitem{Hayes59}
W.~D. Hayes and R.~F. Probstein.
\newblock {\em Hypersonic flow theory}.
\newblock Academic Press, 1959.

\bibitem{Hornung72}
H.~G. Hornung.
\newblock Non-equilibrium flow of nitrogen over spheres and circular cylinders.
\newblock {\em J.~Fluid Mech.}, 53:149--176, 1972.

\bibitem{Hornung2019}
H.~G. Hornung, J.~Martinez Schramm, and K.~Hannemann.
\newblock Hypersonic flow over sphericcally blunted cone capsules for \
  atmospheric entry. {P}art 1, the sharp cone and the sphere.
\newblock {\em J. Fluid Mech.}, 871:1097--1116, 2019.

\bibitem{Pullin80}
D.~I. Pullin.
\newblock Direct simulation methods for compressible inviscid ideal-gas flows.
\newblock {\em J. Comput. Phys}, 34:231--240, 1980.

\bibitem{Quirk94}
J.~J. Quirk.
\newblock A contribution to the great {R}iemann solver debate.
\newblock {\em Int. J. for Num. Methods in Fluids}, 18:555--574, 1994.

\bibitem{Quirk98}
J.~J. Quirk.
\newblock Amrita --- a computational facility (for {CFD} modelling).
\newblock In {\em {VKI} {CFD} Lecture Series}, volume~29. von Karman Institute,
  1998.

\bibitem{QuirkKarni}
J.~J. Quirk and S.~Karni.
\newblock On the dynamics of a shock bubble interaction.
\newblock {\em J. Fluid Mech.}, 318:129--163, 1996.

\bibitem{Stulov69}
V.~P. Stulov.
\newblock Similarity law for supersonic flow past blunt bodies.
\newblock {\em Izv. AN SSSR, Mechanika Zhidkosti i Gaza}, 4:142--146, 1969.

\bibitem{Wen95}
C.-Y. Wen and H.~G. Hornung.
\newblock Non-equilibrium dissociating flow over spheres.
\newblock {\em J.~Fluid Mech.}, 299:389--405, 1995.

\end{thebibliography}

\end{document}